\documentclass[a4paper,11pt]{article}
\pdfoutput=1 % if your are submitting a pdflatex (i.e. if you have
\usepackage{jheppub} % for details on the use of the package, please
\usepackage{slashed}
\usepackage{amsfonts}
\usepackage{amsmath}
\usepackage{amssymb}
\usepackage{bbm}
\usepackage{bbold}
\usepackage[vcentermath]{youngtab}
\usepackage[T1]{fontenc} % if needed
\usepackage{silence}
\usepackage{physics}
\newcommand{\ii}{\mathrm{i}}
\newcommand{\beq}{\begin{eqnarray}}
\newcommand{\eeq}{\end{eqnarray}}

\newcommand{\bmp}{\noindent\begin{minipage}{16cm}}
\newcommand{\emp}{\end{minipage}\vskip 7mm} % 7mm untightened

\newcommand{\SU}{\mbox{SU}}
\newcommand{\SO}{\mbox{SO}}
\newcommand{\SP}{\mbox{Sp}}
\newcommand{\UU}{\mbox{U}}
\newcommand{\st}{s_\theta}
\newcommand{\ct}{c_\theta}

    \newcommand{\fL}{f_{\Lambda}}
    
    \newcommand{\cQ}{c_{Q}}
    \newcommand{\cL}{c_{\Lambda}}

\definecolor{pumpkin}{rgb}{1.0, 0.4, 0.0}

\title{Velocity-dependent self-interacting dark matter and composite Higgs}

\author[a,b]{Martin Rosenlyst}

\affiliation[a]{Rudolf Peierls Centre for Theoretical Physics, University of Oxford, Parks Road, Oxford OX1 3PU, United Kingdom, and}

\affiliation[b]{$ \text{CP}^3$-Origins, University of Southern Denmark,
Campusvej 55, DK-5230 Odense M, Denmark\\}

\emailAdd{mrosenlyst@sdu.dk}

\abstract{We show that the mass of a self-interacting dark matter candidate, specifically a Dirac fermion, can be generated by composite dynamics, with a light scalar mediator emerging alongside the Higgs itself as composite particles. These novel models naturally explain the halo structure problems at various scales and alleviates the Standard Model naturalness problem simultaneously. The relic density of the dark matter candidates is particle anti-particle symmetric and due to thermal freeze-out. These models are four-dimensional gauge theories with a minimal number of fermions charged under a new confining gauge group. Finally, we demonstrate that these models satisfy various constraints set by the dark matter relic density, Big Bang Nucleosynthesis, Cosmic Microwave Background, as well as direct and indirect detection experiments.}

\begin{document} 
\maketitle
\flushbottom

\section{Introduction}

Compelling evidence derived from astronomical and cosmological observations~\cite{Planck:2018vyg} strongly indicates that dark matter (DM) constitutes the predominant form of matter in the Universe, composing approximately $ 85\% $ of its mass. Unraveling the non-baryonic nature of DM stands as one of the most significant mysteries in our comprehension of the physical world. In recent decades, the primary DM paradigm, Weakly Interacting Massive Particles (WIMPs)~\cite{Steigman:1984ac}, faces challenges due to the non-observation of numerous WIMP candidates, notwithstanding advancements in collider, direct-detection and indirect-detection experiments.

Similarly, many inferred DM density profiles near galactic centers and in cluster halos, drawn from observations, exhibit cored (shallow) structures rather than the cuspy (steep) profiles anticipated by N-body simulations of halo formation with WIMPs~\cite{Navarro:1996bv,Navarro:1996gj,Borriello:2000rv,deBlok:2002vgq,954985,2010AdAst2010E}. This inconsistency is known as the core-cusp problem~\cite{Moore:1994yx,1994ApJ,2008ApJ,2011ApJ74220W,2012MNRAS419184A,2012ApJ754L39A,2014ApJ78963A,Oh:2015xoa}, suggesting that significant self-interactions among DM particles themselves, which are known to generate such cores~\cite{Spergel:1999mh,Kamada:2016euw}, could be the dominant interactions among DM particles. Furthermore, simulations featuring only collisionless DM predict a substantial population of massive concentrated subhaloes, contradicting the stellar kinematics of observed satellite galaxies around, for example, the Milky Way and the Andromeda Galaxy~\cite{2011MNRAS,2012MNRAS4221203B,Tollerud:2014zha}, known as the ``too big to fail'' (TBTF) problem. This issue also arises in dwarf galaxies within the Local Group~\cite{2014MNRAS444222G,Kirby:2014sya} and beyond~\cite{Papastergis:2014aba}.

The studies in Refs.~\cite{Spergel:1999mh,Dave:2000ar} and the review in Ref.~\cite{Weinberg:2013aya} demonstrate that the core-cusp and "too big to fail" (TBTF) problems can be alleviated by introducing Self-Interacting Dark Matter (SIDM) particles~\cite{1992ApJ,Hochberg:2014dra} with strong elastic contact-type scattering among themselves, characterized by a velocity-independent cross-section per mass within the range $ \sigma_{SI}/m_{DM}\approx 0.5-1~\text{cm}^2/\text{g} $~\cite{Tulin:2017ara}. However, observations of merging galaxy clusters~\cite{Clowe:2006eq,Randall:2008ppe,2012ApJ747L42D}, halo sphericity~\cite{Peter:2012jh}, and massive clusters~\cite{Kaplinghat:2015aga,2018ApJ853109E,Sagunski:2020spe} suggest that weaker self-interactions are more suitable for these systems, with a cross-section $ \sigma_{\rm SI}/m_{\rm DM}\sim \mathcal{O}(0.1)~\text{cm}^2/\text{g} $. Consequently, these studies advocate for a velocity-dependent cross-section that decreases with increasing scattering velocity, as higher typical relative velocities between DM particles are observed in more massive haloes.

A minimal way to realize this concept is by assuming that the DM can exhibit self-interaction through the exchange of a dark mediator. This SIDM, unlike the collisionless DM in Cold DM (CDM), can engage in elastic scattering among themselves. This characteristic addresses small-scale problems by relying on the velocity dependence of the self-interacting cross section per unit mass, $ \sigma_{\rm SI}/m_{\rm DM} $, typically ranging from $ 0.1-10~\text{cm}^2/\text{g} $ in various small-scale structures~\cite{Tulin:2017ara,Tulin:2013teo}. In this scenario, at least two new particles exist in the hidden sector: the DM particle and the light dark mediator (scalar or vector)~\cite{Bringmann:2020mgx,Dery:2019jwf}.

However, studies have demonstrated that such scenarios are subject to strong constraints, and the simplest considered scenarios, involving DM as a Dirac fermion and a light dark mediator, are usually either excluded or marginally allowed~\cite{Feng:2009hw,Buckley:2009in,Tulin:2013teo,Kaplinghat:2013yxa,DelNobile:2015uua,Bernal:2015ova,Bringmann:2016din,Cirelli:2016rnw,Kahlhoefer:2017umn,Hufnagel:2017dgo,Hufnagel:2018bjp,Bernal:2019uqr}. According to the constraint analysis in Ref.~\cite{Hambye:2019tjt} and first proposed in Ref.~\cite{Chu:2011be}, a potential solution to evade these constraints arises if the DM and light mediator particles have a temperature smaller than the Standard Model (SM) particles.

Remarkably, this situation may naturally occur in the Composite Higgs (CH) framework~\cite{Kaplan:1983fs,Dugan:1984hq}, where two distinct dynamical scales can emerge for the Higgs sector (visible sector) and the DM sector. This development originates from a single strongly interacting extension of the SM. If the portal coupling between these two sectors is weak, the temperature of the DM sector may be smaller than that of the visible sector, thereby expanding the viable parameter space for SIDM models that involve a light mediator. More notably, this hypothesis gains further motivation from the challenges faced by the SM description of the origin of visible mass, specifically the elementary Higgs particle, which is plagued by naturalness and triviality problems. These issues may be alleviated if the Higgs particle itself is composite. These questions are addressed in the CH framework, where the Higgs boson and the light scalar mediator arises as a composite pseudo-Nambu-Goldstone boson (PNGB) from a spontaneously broken global symmetry. Due to this, the Higgs and scalar mediator candidates can be naturally light with respect to the compositeness scale while other composite resonances remain heavy. Moreover, the couplings of the Higgs candidate can be SM-like and therefore these models remain in good agreement with current observations in a large part of the parameter space of the models. 

To almost decouple the two sectors, we will further assume that relevant global symmetries of the Higgs (visible) and DM sectors arise as the chiral symmetries $ G_{1,2} $ of fermions in two different representations of an underlying confining four-dimensional gauge-fermion theory. The strong interactions dynamically break the global symmetries $ G_{1,2} $ to $ H_{1,2} $. Simultaneously, the interactions with the SM fields and an explicit gauge-invariant mass term further break the symmetries, ensuring the correct electroweak symmetry breaking (EWSB) pattern and providing the DM mass, respectively.
Consequently, our investigation explores the potential common composite dynamics origin of the Higgs and SIDM, particularly within CH models. The existing CH research has predominantly focused on the Higgs mechanism and composite DM candidates with weak or no self-interactions.

The paper is organized as follows: In Section~\ref{sec: A minimal composite Higgs model with self-interacting dark matter}, we present
the road to a minimal CH model, including a SIDM candidate with a velocity-dependent self-interacting cross-section, achieved through the exchange of a light composite scalar mediator between the DM particles. In Section~\ref{sec: Dark matter self-interaction}, we explore the viable parameter space of the concrete model, resulting in self-interaction cross sections  consistent with data. In Section~\ref{sec: Dark matter relic density and various constraints}, we investigate the various constraints on the model set by the DM relic density, Big Bang Nucleosynthesis (BBN)
and Cosmic Microwave Background (CMB), as well as direct and indirect detection experiments. In Section~\ref{sec: Model results}, we consider the numerical results of the minimal model. Finally, in Section~\ref{sec: Conclusions}, we give our conclusions.

\section{A minimal composite Higgs model with self-interacting dark matter}
\label{sec: A minimal composite Higgs model with self-interacting dark matter}

This paper presents the first proposal of composite models that predict DM with a velocity-dependent self-interaction cross-section, along with the Higgs itself, as composite pNGBs from a common composite dynamics, allowing for a successful description of the electroweak (EW) symmetry and SIDM. Our proposal features the following key ingredients:  \begin{itemize}
	\item[i)] a SM-like composite pNGB Higgs multiplet with custodial symmetry~\cite{Georgi:1984af}; 
	\item[ii)] a light composite scalar mediator, which achieves a VEV from the new strong interactions, providing the DM mass;
	\item[iii)] a Dirac fermion DM particle which couples to the
light scalar through a Yukawa interaction;
	\item[iv)] strong scattering among DM
particles themselves via the light scalar mediator, with a velocity-dependent cross-section consistent with available data.
\end{itemize} The first ingredient i) is the main feature of CH models~\cite{Kaplan:1983fs}, Little Higgs~\cite{ArkaniHamed:2001nc,ArkaniHamed:2002qx}, holographic extra dimensions~\cite{Contino:2003ve,Hosotani:2005nz}, Twin Higgs~\cite{Chacko:2005pe} and elementary Goldstone Higgs models~\cite{Alanne:2014kea}. In the CH framework, the aim of a composite scalar, weakly interacting with the visible sector, has been accommodated by extending the global symmetry, typically producing thermal DM candidates~\cite{Ma:2015gra,Ma:2015gra,Cai:2018tet,Alanne:2018xli,Cacciapaglia:2019ixa}. Even in Ref.~\cite{Rosenlyst:2021elz}, a light composite scalar is demonstrated to strongly self-interact among themselves; however, it exhibits a velocity-independent cross section. Therefore, until now, none of these model types satisfies ii)-iv).

To realize this, we propose for concreteness four-dimensional gauge theories with a single strongly interacting gauge group $G_{\rm HC}$, with $N_1$ Weyl fermions (denoted $ Q_1 $) in the representation $\mathcal{R}_1$ and $N_2$ Weyl fermions ($ Q_2 $) in the representation $\mathcal{R}_2$. The $\mathcal{R}_1$ fermions are gauged under the EW interactions, while the $R_2$ fermions are inert under the SM gauge symmetry for simplicity.~\footnote{The $R_2$ fermions may be charged under some new gauge symmetries, for example, an $ \SU(2)_{\rm R} $ gauge symmetry as in Ref.~\cite{Cacciapaglia:2021aex}.} Upon the compositeness scale, the $ \mathcal{R}_1 $ fermions form the composite Higgs doublet with the decay constant $ f_{Q_1} $ accommodating i), while the $ \mathcal{R}_2 $ fermions with $ f_{Q_2} $ result in composite light scalar mediator charged under a global $ \UU(1) $ symmetry. Explicitly breaking this symmetry through the introduction of vector-like masses for the fermions $ Q_2 $ leads to a VEV of the scalar mediator, satisfying ii). However, it may be possible to obtain a light scalar mediator candidate consisting of $ \mathcal{R}_2 $ fermions charged under the SM gauge symmetry, but for such a model, it turns out to be challenging to implement the novel mechanism that satisfy various constraints. 
%Furthermore, there exists a global $\UU(1)_\Theta$ symmetry under which both sets of hyper-fermions are charged, which is spontaneously broken by the condensates and generates a singlet pNGB, $ \Theta $, with the decay constant $ f_\Theta $. The kinetic term of $ \Theta $ is canonically normalized if $ f_{Q_1}=f_{Q_2}=f_\Theta\equiv f $, which we will assume for simplicity. Generally, the kinetic terms must be renormalized, but we expect them to be of the same size based on Casimir scaling~\cite{Ryttov:2008xe,Frandsen:2011kt,Alanne:2018xli}. 
Furthermore, we add a new class of interactions that both dynamically induce SM-fermion masses and align the vacuum of the $ \mathcal{R}_1 $ fermions into the CH regime. We will discuss fermion partial compositeness (PC) type~\cite{Kaplan:1991dc} as an example of such interactions. In that case, we add a third sector with QCD charged fermions to accommodate top partners. Finally, we introduce a four-fermion interaction between the Dirac fermion DM candidate $ X $ and the hyper-fermions $ Q_2 $. This interaction results in a Yukawa interaction between $ X $ and $ \Phi $ upon condensation, providing the ingredient iii) and producing the DM mass. To assure condition iv) and generate large enough self-interaction cross section with the right velocity-dependence, the DM candidate and the scalar mediator require, respectively, masses of $ m_{DM}\lesssim 1.5~\text{GeV}$ and $ m_\Phi < 2m_e \approx 1 $~MeV (set by BNN constraints~\cite{Hambye:2019tjt}), which are much smaller than the Higgs decay constant of $ f \gtrsim \mathcal{O}(1)~\text{TeV}$. These small masses can be obtained by having small vector-like masses of hyper-fermions $ Q_2 $. 

%This lightness of these states entail no fulfillment of condition iii) due to kinematics since the remaining composite states, with typical masses of about $\mathcal{O}(f) $ (except for the Higgs boson), are much heavier than these DM candidates. However, in the models considered, there typically exists a pNGB singlet, $ \eta $, in the sector of the $ \mathcal{R}_1 $ hyper-fermions that can mix with the pNGB singlet $ \Theta $ by supposing that only one $ \mathcal{R}_1 $ hyper-fermion doublet achieves an explicit vector-like mass. In that case, the two pNGBs $ \Theta $ and $ \eta $ will almost mix half-and-half, leading to a very light singlet consisting mostly of $ \Theta $ along the lines of Refs.~\cite{Ferretti:2016upr,Belyaev:2016ftv}. It may allow the DM candidates to annihilate into these light singlets and thermally produce the relic density. We, therefore, need an efficient annihilation mode of the DM candidates into the light $ \Theta $. 

\begin{table}[t]
    \begin{center}
	\begin{tabular}{ccccc}
%	    \toprule
	    \hline
	    & $\vphantom{\frac{\frac12}{\frac12}}\quad G_{\mathrm{HC}}\quad$ & $\quad\SU(2)_{\mathrm{W}}\quad$ 
    & $\quad\mathrm{U}(1)_{\mathrm{Y}}\quad$ & $\quad\mathrm{U}(1)_{\Lambda}\quad$ \\
%	    \colrule
	    \hline
	    $\vphantom{\frac{\frac12}{\frac12}} (U_L,D_L)$	&   ${\tiny \yng(1)}$	&   ${\tiny \yng(1)}$	&   0 & 0\\    
	    $\vphantom{\frac{1}{\frac12}} \widetilde{U}_{\mathrm{L}}$	&   ${\tiny \yng(1)}$	&   1	&   $-1/2$ & 0\\    
	    $\vphantom{\frac{1}{\frac12}} \widetilde{D}_{\mathrm{L}}$	&   ${\tiny \yng(1)}$	&   1	&   $+1/2$  & 0
	    \\  
	   $\vphantom{\frac{1}{\frac12}} \lambda_{ \mathrm{L}}$	&   ${\rm Adj}$	&   1	&   $0$ & $ +1/2 $ \\    
	   $\vphantom{\frac{1}{\frac12}} \widetilde{\lambda}_{ \mathrm{L}}$	&   ${\rm Adj}$	&   1	&   $0$ & $ -1/2 $ \\  
%	    \botrule
	    \hline
	\end{tabular}
    \end{center}
        \caption{The new fermion content and their charges, as in Refs.~\cite{Ryttov:2008xe,Alanne:2018xli}. All groups are gauged except for
$ \UU(1)_\Lambda $, which is a global symmetry in the $ \Lambda $--sector responsible for dark matter stability.}
    \label{table:fullmodel}
\end{table}

For concreteness, for one of the hyper-fermion sectors with a single representation, the symmetry breaking patterns are known~\cite{Witten:1983tx,Kosower:1984aw}: Given $N$ Weyl fermions transforming as the $\mathcal{R}$ representation of $ G_{\rm HC} $, the three possible classes of vacuum cosets are $ \SU(N)/\SO(N) $ for real $\mathcal{R}$, $ \SU(N)/\SP(N) $ for pseudo-real $\mathcal{R}$ and $ \SU(N)\otimes \SU(N)\otimes \UU(1)/ \SU(N)\otimes \UU(1)$ for complex $\mathcal{R}$~\cite{Peskin:1980gc}. For these three classes, the minimal CH cosets that fulfill requirement i) contain $N=5$ in the real case~\cite{Dugan:1984hq}, and $N=4$ in both the pseudo-real~\cite{Galloway:2010bp} and the complex cases~\cite{Ma:2015gra}. In terms of pNGB spectrum, the pseudo-real case is the most minimal, with only five states, satisfying condition i). Furthermore, the minimal DM cosets that can fulfill the condition ii) contain $N=2$ in the real~\cite{Ryttov:2008xe}, $N=4$ in the pseudo-real~\cite{Ryttov:2008xe} and $N=2$ in the complex case~\cite{Kaplan:1983fs}. We will, therefore, focus on the real case since it only needs two Weyl fermions and has the most minimal pNGB coset, with one complex pNGB singlet playing the role of light scalar mediator. In the following, these two minimal subsets will be combined in one minimal model example, fulfilling all the requirements i)-iv).

\subsection{The underlying Lagrangian}

From now on, we focus on the minimal model example that fulfills all the conditions i)-iv). In this minimal model example, the gauge group can be chosen to be the minimal $G_{\mathrm{HC}}=\SU(2)_{\mathrm{HC}}$ with $\mathcal{R}_1$ as the fundamental representation and $N_1=4$ Weyl spinors, arranged in one $\SU(2)_{\rm L}$ doublet $(U_L,D_L)$ and two singlets $\widetilde{U}_L$ and $ \widetilde{D}_L$. We further take $\mathcal{R}_2$ to be the adjoint representation and $N_2=2$ Weyl spinors, arranged in two SM singlets $\lambda_L$ and $ \widetilde{\lambda}_L $, as in Refs.~\cite{Ryttov:2008xe,Alanne:2018xli}. The fermion content in terms of left-handed Weyl fields, with
$ \widetilde{\psi}_{\mathrm{L}}\equiv \epsilon \psi_R^* $, along with their EW quantum numbers are presented in Table~\ref{table:fullmodel}. Finally, the spinors $Q=(U_{\mathrm{L}},\, D_{\mathrm{L}},\, \widetilde{U}_{\mathrm{L}},\, \widetilde{D}_{\mathrm{L}})$ 
and $\Lambda=(\lambda_{ \mathrm{L}}, \,   \widetilde{\lambda}_{ \mathrm{L}})$
transform in the fundamental representations of the $\SU(4)_Q$ and $\SU(2)_\Lambda$ subgroups of the global symmetry, respectively, where we drop an L subscript on $Q$ and $\Lambda$ for simplicity. 

In terms of $Q$ and $\Lambda$, the underlying gauge-fermion Lagrangian of the CH model can be written as \beq 
\mathcal{L}_{\rm CH}= Q^\dagger i \gamma^\mu D_\mu Q+\Lambda^\dagger i \gamma^\mu D_\mu \Lambda - \frac{1}{2} \left(Q^T M_Q Q +\Lambda^T M_\Lambda \Lambda +{\rm h.c.} \right)
	 + \mathcal{L}_{ 4 f}\,, 
\label{eq: Basic Lagrangian (UV)}
\eeq where the covariant derivatives include the HC--gluons and the 
$\SU(2)_{\rm L}$ and $ \UU(1)_{\rm Y} $ gauge bosons. The mass matrices of the hyper-fermions $ Q $ and $ \Lambda $ are, respectively, given by \begin{equation} \label{eq: hyper-fermion masses}
    M_Q = \left( \begin{array}{cc}
	m_1 \ii \sigma_2 & 0 \\
	0 & -m_2\ii \sigma_2
    \end{array} \right)\,,\quad\quad
 M_{\Lambda}=\left( \begin{array}{cc}
	-\mu & m \\
	m & \mu
    \end{array} \right)\,, 
\end{equation} where $ \sigma_2 $ is the second Pauli matrix. 
The terms in $ \mathcal{L}_{4f}$ in Eq.~(\ref{eq: Basic Lagrangian (UV)}) are four-fermion interactions responsible for generating the masses and Yukawa couplings for the SM fermions and the DM fermion in the condensed phase. 

%The mass term contributes to the correct vacuum alignment in between the EW unbroken vacuum and a Technicolor vacuum, and consists of two independent masses $ \overline{m}_{1,2} $ for the doublet and singlet hyper-fermions, $M_Q=\text{Diag}(i\overline{m}_1\sigma_2,-i\overline{m}_2\sigma_2)$, where $ \sigma_2 $ is the second Pauli matrix. 

%The remaining essential ingredient for model building is the inclusion of operators responsible for generating SM fermion masses and aligning the vacuum of the $ Q $-sector.

For the hyper-fermion $ Q $, these four-fermion operators can be written as PC-type~\cite{Kaplan:1991dc}. These require the extension of the model in Table~\ref{table:fullmodel} by a new species of fermions $\chi_t$, transforming under the two-index anti-symmetric representation of $ G_{\rm HC}=\SP(2N)_{\rm HC} $, and carrying appropriate quantum numbers under the SM gauge symmetry. For the top, it is enough to introduce a $\SU(2)_{\rm W}$ vector-like pair with hypercharge $+2/3$ and transforming as a fundamental of the ordinary $ \SU(3)_{\rm C} $ color gauge group (QCD). Models of this type were first proposed in Refs.~\cite{Barnard:2013zea,Ferretti:2013kya} and our model is an extension of the one in Ref.~\cite{Barnard:2013zea}. We arrange these new QCD colored hyper-fermions into an $ \SU(6)_\chi $ vector, which will spontaneously break to $ \SO(6)_\chi $ upon the condensation. 

Thus, we study the minimal CH model with global symmetry breaking pattern \begin{equation}
    \begin{split}
\frac{\SU(4)_Q \otimes \SU(2)_\Lambda\otimes \SU(6)_\chi \otimes \UU(1)_{\Theta_1}\otimes \UU(1)_{\Theta_2} }{\SP(4)_Q\otimes \UU(1)_{\Lambda}\otimes \mathbb{Z}_2\otimes \SO(6)_\chi} \,.
    \end{split}
\label{eq:NGBmatrixMod}
\end{equation} Now, there are two anomaly-free U(1) corresponding to two pNGBs, $ \Theta_{1,2} $, and one anomalous U(1) corresponding to the state $ \Theta' $. 
The $Q$, $\Lambda$ and $\chi$ charges under these U(1) are defined by the anomaly cancellation
\begin{equation}
q_Q T(\mathbf{F}) + q_\Lambda T(\mathbf{G}) +q_\chi T(\mathbf{A_2})=0\Rightarrow q_\Lambda = - \frac{1}{3}q_Q-q_\chi
\end{equation} with $T(\mathcal{R})=1,\,2N_{\mathrm{HC}}+2,\,2N_{\mathrm{HC}}-2$ the index of representation $\mathcal{R}={\bf F,\,G,\,A_2}$, respectively. In our case, we have $n=3$ fermionic sectors, $ Q $ in {\bf F}, $\Lambda$ in {\bf G} and $\chi$ in $\bf A_2$ of $G_{\mathrm{HC}}=\mathrm{Sp}(4)_{\mathrm{HC}}$. Therefore, the anomalous charges are $q_{F,3}=1$, $q_{G,3}=3$, $q_{A_2,3}=3$ and the charges of the two anomaly-free are the orthogonal combinations $ q_{F,1}=-3 $, $ q_{G,1}=1 $, $ q_{A_2,1}=0 $ and $ q_{F,2}=0 $, $ q_{G,2}=1 $, $ q_{A_2,2}=-1 $. Moreover, the QCD gauge group $ \SU(3)_{\rm C} $ will be identified as a subgroup of the unbroken group $ \SO(6)_\chi $. This global symmetry breaking pattern can thus determine the possible hypercolor groups~\cite{Ferretti:2013kya}, which can only be $ G_{\rm HC}=\SP(2N)_{\rm HC} $ with $ 2\leq N \leq 18 $ (or $ G_{\rm HC}=\SO(N)_{\rm HC}$ with $ N=11,13 $ for $ Q $ and $ \chi $ transforming, respectively, in the spin and fundamental representations under $ G_{\rm HC} $). Therefore, the minimal choice is $ G_{\rm HC}=\SP(4)_{\rm HC} $. 

%$ \SU(4)_Q /\SP(4)_Q \otimes \SU(6)_\chi/\SO(6)_\chi  \otimes [\UU(1)_Q \otimes \UU(1)_\chi]/\varnothing  $

The four-fermion interactions that will generate the PC operators we consider are~\cite{Alanne:2018wtp}: 
\beq \label{eq: top PC-operators}
 \frac{\widetilde{y}_{L}}{\Lambda_t^2} q^{\alpha\dagger}_{L,3} (Q^\dagger P_q^\alpha Q^*\chi_{t}^\dagger) + \frac{\widetilde{y}_{R}}{\Lambda_t^2}t_{R}^{c\dagger} (Q^\dagger P_t Q^* \chi_{t}^\dagger)+\rm h.c.\,, 
\eeq  
where $ q_{L,3} $ and $ t_R $ are the third generation of the quark doublets and the top singlet, respectively, and the spurions that project onto the appropriate components in the fourplet $Q$ are given by \begin{equation}
    P_q^1 =\frac{1}{\sqrt{2}} \left( \begin{array}{cccc}
	0 & 0 & 1 & 0 \\
	0 & 0 & 0 & 0 \\
	1 & 0 & 0 & 0 \\
	0 & 0 & 0 & 0 
    \end{array} \right)\,,\quad
  P_q^2 =\frac{1}{\sqrt{2}} \left( \begin{array}{cccc}
	0 & 0 & 0 & 0 \\
	0 & 0 & 1 & 0 \\
	0 & 1 & 0 & 0 \\
	0 & 0 & 0 & 0 
    \end{array} \right)\,,\quad 
    P_t =\frac{1}{\sqrt{2}} \left( \begin{array}{cccc}
	0 & 0 & 0 & 0 \\
	0 & 0 & 0 & 0 \\
	0 & 0 & 0 & 1 \\
	0 & 0 & 1 & 0 
    \end{array} \right)
\end{equation} For both the left- and right-handed top, we choose the spurions to transform as the two-index symmetric of the unbroken chiral symmetry subgroup $ \SU(4)_Q $ as the minimal possible choice. 

Finally, the four-fermion operator in $ \mathcal{L}_{4f}$ in Eq.~(\ref{eq: Basic Lagrangian (UV)}) responsible for generating the mass and scalar-mediator Yukawa coupling for the DM fermion in the condensed phase is given by \beq \label{eq: four-fermion operator}
 \frac{\widetilde{y}_{\Phi}}{\Lambda_\Phi^2}(X_L X_{R}^{c})^\dagger (\Lambda^T P_\Phi \Lambda)+\rm h.c.\quad \text{with}\quad    P_\Phi =\frac{1}{2} \left( \begin{array}{cc}
	1 & 0\\
	0 & 0 
    \end{array} \right)\,.
\eeq 

\subsection{The electroweak embedding and the condensates}

Upon the condensation of the hyper-fermions at the scale $ \Lambda_{\rm HC}\sim 4\pi f $, with $ f $ as the decay constant of the composite pNGBs, they can form an anti-symmetric and SM gauge invariant condensate of the form \beq
 \langle Q^I_{\alpha,a}Q^J_{\beta,b}\rangle\epsilon^{\alpha\beta}\epsilon^{ab}\sim f_Q^3 E_Q^{IJ} \,, \quad  \langle \Lambda^A_{\alpha,i}\Lambda^B_{\beta,j}\rangle\epsilon^{\alpha\beta}\epsilon^{ij}\sim f_\Lambda^3 E_\Lambda^{AB} \,,
\eeq
where $ \alpha,\beta $ are spinor indices, $ a,b,i,j $ are HC indices, and $ I,J,A,B $ are flavour indices of the hyper-fermions. The condensates break the global symmetries $ \SU(4)_Q $ to $ \SP(4)_Q $ and $ \SU(2)_\Lambda \otimes \UU(1)_{\Theta_1} $ to $ \UU(1)_\Lambda\otimes \mathbb{Z}_2 $, respectively. %We anticipate the Goldstone boson decay constants, denoted as $ f_Q $ and $ f_\Lambda $, to be of comparable magnitudes~\cite{Ryttov:2008xe,Frandsen:2011kt}. For simplicity, we will assume their identity later on. 
$ f_{Q,\Lambda} $ are the decay constants of the composite pNGBs in the $ Q $ and $ \Lambda $ sector, respectively. In many scenarios, the decay constants $ f_{Q,\Lambda} $ of the composite pNGBs in the $ Q $ and $ \Lambda $ sector, respectively, are anticipated to be of comparable magnitudes~\cite{Ryttov:2008xe,Frandsen:2011kt}, i.e. $ f\equiv f_Q \sim f_\Lambda $. However, as we will consider later, these decay constants can differ by several orders of magnitude, as shown in Ref.~\cite{Frandsen:2011kt}.

To describe the general vacuum alignment in the effective Lagrangian, we identify an $\SU(2)_{\rm L}\otimes\SU(2)_{\rm R}$ subgroup in $\SU(4)_Q$ by the left and right generators
\begin{equation}
    \label{eq:gensLR}
    T^i_{\mathrm{L}}=\frac{1}{2}\left(\begin{array}{cc}\sigma_i & 0 \\ 0 & 0\end{array}\right)\,, \quad \quad
    T^i_{\mathrm{R}}=\frac{1}{2}\left(\begin{array}{cc} 0 & 0 \\ 0 & -\sigma_i^{T}\end{array}\right)\,,
\end{equation}
where $\sigma_i$ are the Pauli matrices.  The EW subgroup is gauged after identifying the generator of 
hypercharge with $T_{\mathrm{R}}^3$. The alignment between the EW subgroup and the stability group $\SP(4)_Q $ can then be conveniently parameterized by an angle, $\theta$, after identifying the vacua that leave the 
EW symmetry intact, $E_Q^{\pm}$, and the one breaking it completely to $\mathrm{U}(1)_{\mathrm{EM}}$ 
of electromagnetism, $E_Q^{\mathrm{B}}$, with:
\begin{equation}
    E_{Q}^{\pm} = \left( \begin{array}{cc}
	\ii \sigma_2 & 0 \\
	0 & \pm \ii \sigma_2
    \end{array} \right)\,,\quad\quad
    E_{Q}^B  =\left( \begin{array}{cc}
	0 & \mathbb{1}_2 \\
	-\mathbb{1}_2 & 0
    \end{array} \right) \,,\quad \quad
 E_{\Lambda}=\left( \begin{array}{cc}
	0 & 1 \\
	1 & 0
    \end{array} \right)\,,
\end{equation}
where we have also written the $\Lambda$--sector vacuum matrix, $E_{\Lambda}$.  
The true $\SU(4)_Q$ vacuum can be written as a linear combination of the EW-preserving and EW-breaking vacua, 
${E_Q=c_\theta E_Q^-+s_\theta E_Q^{\mathrm{B}}}$. We use the short-hand notations $s_x\equiv \sin x, c_x\equiv \cos x$, and $t_x\equiv \tan x$ throughout.
Either choice of $E_Q^{\pm}$ is equivalent~\cite{Galloway:2010bp}, and in this paper, we have chosen $E_Q^-$. 

The Goldstone excitations around the vacua are then parameterized by  
\begin{equation}
    \begin{split}
    \Sigma_Q &= \exp\left[ 2\sqrt{2}\, i  \left(\frac{\Pi_Q}{f}-\frac{3}{2\sqrt{19}}\frac{\Theta_1}{f_{\Theta_1}}\mathbb{1}_4\right) \right] E_Q\,, \\
    \Sigma_\Lambda &= \exp\left[ 2\sqrt{2}\,i \left( \frac{\Pi_\Lambda}{\fL}+\frac{1}{2\sqrt{19}}\frac{\Theta_1}{f_{\Theta_1}} \mathbb{1}_2+\frac{1}{2}\frac{\Theta_2}{f_{\Theta_2}} \mathbb{1}_2\right)\right] 
    E_\Lambda
    \end{split}
\label{eq:NGBmatrixMod}
\end{equation} with
\begin{equation}
    \Pi_Q=\sum_{i=1}^5 \Pi_Q^ iX^i_Q , \quad\quad  \Pi_\Lambda=\sum_{a=1}^2 \Pi_\Lambda^a X^a_\Lambda\,,
\end{equation}
where $X_{Q,\Lambda}$ are the $\theta$-dependent broken generators of $ \SU(4)_Q $ and $ \SU(2)_\Lambda $ and can be found explicitly in Refs.~\cite{Ryttov:2008xe,Galloway:2010bp}. Here the $ \Theta_2 $ state in the $ \chi $--sector is typically decoupled by adding an explicit mass, $ m_\chi $. 
The $\Theta_1$ state is the only state connecting the $ Q $ and $ \Lambda $ sectors in the effective Lagrangian. 
%The NGB matrices in Eq.~(\ref{eq:NGBmatrix}) will be modified to Eq.~(\ref{eq:NGBmatrixMod}) when we add the PC fermion mass operators in Eq.~(2). 
In the composite Higgs range, $ 0<\theta < \pi/2 $, we identify $h\equiv\Pi_Q^4 \sim \ct (\bar{U}U+\bar{D}D) + \st \,{\rm Re} \, U^TCD $  as the composite Higgs and $\eta\equiv \Pi_Q^5 \sim \,{\rm Im} \, U^TCD  $, while the remaining three $ \Pi_L^{1,2,3} $ are exact Goldstones eaten by the massive $ W^\pm $ and $ Z $. Furthermore, we identify the scalar mediator as  $\Phi\equiv\frac{1}{\sqrt{2}}(\Pi_\Lambda^1-i \Pi_{\Lambda}^2)\sim \Lambda^T C \Lambda$ and $\bar{\Phi}\equiv\frac{1}{\sqrt{2}}(\Pi_\Lambda^1+i \Pi_{\Lambda}^2)\sim \overline{\Lambda}^T C \overline{\Lambda} $ in the $\Lambda$--sector, while $ \Theta_1 \sim i (\overline{U}\gamma^5 U+\overline{D}\gamma^5 D-(1/2)\overline{\Lambda}\gamma^5 \Lambda) $. Note that, following Ref.~\cite{Alanne:2018xli}, we have here used Dirac spinors to indicate the hyper-fermions. 

\subsection{Effective Lagrangian and vacuum alignment}

Upon the compositeness scale, $ \Lambda_{\rm HC}\sim 4\pi f $, the effective Lagrangian can be written as
\begin{equation}
    \label{eq:effLag}
    \mathcal{L}_\mathrm{eff}=\mathcal{L}_{\mathrm{kin}}+\mathcal{L}_{\mathrm{f}}-V_{\mathrm{eff}}\,,
\end{equation}
where the kinetic terms are
\begin{equation}
    \label{eq:kinLag}
    \mathcal{L}_{\mathrm{kin}}=\frac{f^2}{8}\Tr [D_{\mu}\Sigma_Q^{\dagger}D^{\mu}\Sigma_Q]+\frac{\fL^2}{8}\Tr [\partial_{\mu}\Sigma_\Lambda^{\dagger}\partial^{\mu}\Sigma_\Lambda]
\end{equation}
with 
\begin{equation}
    \label{eq:covD}
    D_{\mu}\Sigma_Q=\partial_{\mu}\Sigma_Q-\ii\left(G_{\mu}\Sigma_Q+\Sigma_QG_{\mu}^{T}\right)\,,
\end{equation}
and the EW gauge fields are encoded in the covariant derivative
\begin{equation}
    \label{eq:Gmu}
    G_{\mu}=g_L W_{\mu}^iT_{\mathrm{L}}^i+g_Y B_{\mu}T_{\mathrm{R}}^3\,.
\end{equation} Besides providing kinetic terms and self-interactions for the pNGBs, it will induce masses for the EW gauge bosons and their couplings with the pNGBs (including the SM Higgs identified as $ h $), 
\beq \label{WZ masses and SM VEV}
&&m_W^2=\frac{1}{4}g_L^2f^2s_\theta^2\,,\quad\quad m_Z^2=m_W^2/c^2_{\theta_W}\,, \\
&&g_{hWW}=\frac{1}{2}g^2_Wfs_\theta c_\theta=g_{  hWW}^{\rm SM}c_\theta\,,\quad\quad g_{ hZZ}=g_{ hWW}/c^2_{\theta_W}\,, \nonumber
\eeq 
where $ v_{\rm SM}\equiv fs_\theta = 246~\text{GeV} $, $ g_{L,Y} $ are the EW $ \SU(2)_{\rm L} $ and $ \UU(1)_{\rm Y} $ gauge couplings, and $ \theta_W $ is the Weinberg angle. 
The vacuum misalignment angle $ \theta $ parametrizes the deviations of the CH Higgs couplings to the EW gauge bosons with respect to the SM Higgs. These deviations are constrained by direct LHC measurements~\cite{deBlas:2018tjm} of this coupling which imply an upper bound of $ s_\theta \lesssim 0.3 $. EW precision measurements also impose an upper limit which has been found to be stricter $ s_\theta \lesssim 0.2 $~\cite{Cacciapaglia:2020kgq}. However, in Ref.~\cite{Frandsen:2022xsz}, it is shown that the constraints on $ s_\theta $ from EW precision measurements are alleviated when all contributions from the composite fermion resonances in the CH models are included.

The PC four-fermion operators in Eq.~(\ref{eq: top PC-operators})  contribute to $ \mathcal{L}_{\mathrm{f}} $ in Eq.~(\ref{eq:effLag}) with effective operators generating the top mass and Yukawa coupling, which can be written as~\cite{Alanne:2018wtp}
\begin{equation}
    \begin{split} \label{eq: top Yukawa operator}
     \mathcal{L}_{\mathrm{f}}&\supset \frac{C_{yS}}{4\pi} y_{L} y_{R}f (t_L t^c_R)^\dagger~{\rm Tr} [P_Q^1 \Sigma^\dagger_Q] {\rm Tr} [P_\Lambda\Sigma_\Lambda^\dagger] +\mathrm{h.c}\\
    &=(t_L t^c_R)^\dagger \left( m_\mathrm{top} + \frac{m_\mathrm{top}}{v_{\rm SM}} \left(c_\theta \frac{c_{2\theta}}{c_\theta} h-i\frac{s_\theta}{c_\theta}\eta \right)
	 + \dots \right) +\mathrm{h.c.}\,, 
    \end{split}
\end{equation}
where $C_{yS}\sim\mathcal{O}(1)$,  $m_{\rm top} =C_{yS} y_{L} y_{R} v_{\rm SM} /(2\pi)$, and $ y_{L/R} $ are related to the couplings $ \widetilde{y}_{L/R} $ via the anomalous dimensions of the fermionic operators and are expected to be $ \mathcal{O}(1) $ for the top quark.

The $ \mu $-mass term in Eq.~(\ref{eq: Basic Lagrangian (UV)}) breaks the $ \UU(1)_\Lambda $ symmetry, resulting in a VEV for $ \Phi $. From now on, we adopt the linear parametrization for $ \Phi $. We can express
it as \begin{equation}
    \label{eq:linear parametrization}
 \Phi=\frac{1}{\sqrt{2}}(v_\Phi + \phi_R +i \phi_I) \,.
\end{equation} Thus, the four-fermion operator in Eq.~(\ref{eq: four-fermion operator}) contributes to the DM mass and Yukawa interaction in $ \mathcal{L}_{\mathrm{f}} $ in Eq.~(\ref{eq:effLag}), which yields~\cite{Alanne:2018wtp} \begin{equation}
\begin{split} \label{eq: DM mass and Yukawa}
\mathcal{L}_{\mathrm{f}}&\supset y_\Phi f_\Lambda(X_L X_{R}^{c})^\dagger {\rm Tr}[P_\Phi \Sigma_X] +\mathrm{h.c.}\\
    &=(X_L X^c_R)^\dagger \left(m_X + y_\Phi \Phi 
	 + \dots \right) +\mathrm{h.c.}
    \end{split}
\end{equation} with $ y_\Phi\equiv (4\pi)^3 N_{\rm HC} A (f_\Lambda / \Lambda_\Phi)\widetilde{y}_{\Phi} $~\cite{Hill:2002ap}, where $ N_{\rm HC} $ is the number of hypercolors and $ A $ is an integration
constant arising from the condensation. Therefore, the DM mass is $ m_X = y_\Phi v_\Phi /\sqrt{2}  $.

The value of $\theta$, and the amount of misalignment, is controlled by 
the effective potential  $V_{\mathrm{eff}}$ in Eq.~(\ref{eq:effLag}), which receives contributions from
 the EW gauge interactions, the vector-like masses of the hyper-fermions and the SM fermion couplings 
to the strong sector. At leading order, each source of symmetry breaking contributes independently to
the effective potential:
\beq
V_{\text{eff}}&=&V_{\text{gauge}}+V_\text{m} +V_{\text{top}}+V_{X}+\dotsc \,, \label{Potential 1}
\eeq 
where the dots are left to indicate the presence of mixed terms at higher orders,
or the effect of additional UV operators.
In this work, we will write the effective potential in terms of effective operators,
which contain insertions of spurions that correspond to the symmetry breaking 
couplings. A complete classification of such operators, for this kind of cosets, up
to next-to-leading order can be found in Ref.~\cite{Alanne:2018wtp}.

Both the contribution of gauge interactions and of the hyper-fermion masses arise
to leading $\mathcal{O} (p^2)$ order and have a standard form:
\begin{equation}
\begin{aligned} V_{\rm gauge,p^2}&=- C_g f^4 \left( \sum_{i=1}^3 g_L^2 \text{Tr}[T_L^i \Sigma_Q T_L^{iT}\Sigma_Q^\dagger]+g_Y^{2}\text{Tr}[T_R^3 \Sigma_Q T_R^{3T} \Sigma_Q^\dagger]\right) \,, \end{aligned} \end{equation} 
where the non-perturbative $ \mathcal{O}(1) $ coefficient $ C_g $ can be determined by the lattice simulations~\cite{Arthur:2016dir} and $T_{L/R}$ are given in Eq.~(\ref{eq:gensLR}), while \begin{equation}
\begin{aligned}
    \label{eq:V0}
   V_{\rm m,p^2}&=2\pi\cQ f^3\,\Tr\left[M_Q\Sigma_Q^{\dagger}+\Sigma_QM_Q^{\dagger}\right]
    +2\pi\cL \fL^3\,\Tr [M_{\Lambda}\Sigma_{\Lambda}^{\dagger}+\Sigma_{\Lambda}M_{\Lambda}^{\dagger}]\,,
\end{aligned}\end{equation} where the coefficients $\cQ$ and $\cL$ are non-perturbative $\mathcal{O}(1)$ constants, and we use the numerical value ${\cQ\approx 1.5}$ suggested by the lattice simulations~\cite{Arthur:2016dir}. The mass terms involving $M_Q$ (as well as the subleading EW gauge interactions) prefer the vacuum where the EW is unbroken. The correct vacuum alignment must, therefore, be ensured by the SM-fermion mass generation mechanism. The top loop contributions arising from the PC four-fermion operator in Eq.~(\ref{eq: top PC-operators}) yield at leading $\mathcal{O} (p^4)$ order in the chiral expansion the effective potential contribution as follows~\cite{Alanne:2018wtp}  \begin{equation}
\begin{split} \label{eq: LO terms}
V_{\rm top,p^4}=&\frac{C_{tS}}{(4\pi)^2}f^4\bigg( y_L^4{\rm Tr}[P_Q^\alpha \Sigma^\dagger P_Q^\beta \Sigma^\dagger]{\rm Tr}[\Sigma P_{Q\alpha}^\dagger \Sigma P_{Q\beta}^\dagger]+y_R^4 {\rm Tr}[P_t \Sigma^\dagger P_t \Sigma^\dagger ]{\rm Tr}[\Sigma P_t^\dagger \Sigma P_t^\dagger]\\ &\phantom{\frac{C_{tS}}{(4\pi)^2}f^4\bigg(}+y_L^2 y_R^2 {\rm Tr}[P_Q^\alpha \Sigma^\dagger P_t \Sigma^\dagger ] {\rm Tr}[\Sigma P_{Q\alpha}^\dagger \Sigma P_t^\dagger]\bigg)\,,
    \end{split}
\end{equation} which contributes to the total effective Lagrangian $ V_{\rm eff} $ in Eq.~(\ref{eq:effLag}). Here $ C_{tS} $ is a non-perturbative $ \mathcal{O}(1) $ constant.  

The DM loop contributions arising from the four-fermion operator in Eq.~(\ref{eq: four-fermion operator}) yield at leading $\mathcal{O} (p^2)$ order in the chiral expansion the effective potential contribution as follows~\cite{Alanne:2018wtp} \begin{equation}
\begin{split} \label{eq: LO terms}
V_{\rm X,p^2}=&-C_X y_\Phi^2 f_\Lambda^4 {\rm Tr}[P_X \Sigma_\Lambda]^2\,,
    \end{split}
\end{equation} where $ C_{X} $ is a non-perturbative $ \mathcal{O}(1) $ constant.

%Finally, it is relevant for the DM mass to consider possible four-hyper-fermion operators in the $\Lambda$--sector of the form $ \Lambda \Lambda \Lambda \Lambda $, which yield potential contributions~\cite{Foadi:2007ue}:\begin{align} V_{\Lambda}&=C_\Lambda g_\Lambda^2 f^4_\Lambda  \Tr[\sigma_3\Sigma_\Lambda^{\dagger}\sigma_3\Sigma_\Lambda]\,,\label{eq: Lambda4 term} \end{align} which preserves the symmetry $ \mathrm{U}(1)_\Lambda $ and has no effects on the vacuum alignment. The coefficient $ C_\Lambda $ is $ \mathcal{O}(1) $ form factor that can be computed on the lattice~\cite{Arthur:2016ozw}.

Finally, from the single trace terms in Eq.~(\ref{Potential 1}), the only interactions between the $ \mathcal{R}_1 $ and $ \mathcal{R}_2 $
sectors are those involving the $ \Theta_1 $ state. At the next leading order, all interactions between the $ \mathcal{R}_1 $ and $ \mathcal{R}_2 $ sectors arise from double trace terms, which are given by \begin{equation}
\begin{split} \label{eq: double trace terms}
\mathcal{L}_{Q,\Lambda}=&\frac{c_1}{4\pi}{\rm Tr}[D_\mu \Sigma_Q^\dagger D^\mu \Sigma_Q]{\rm Tr}[\partial_\mu \Sigma_\Lambda^\dagger \partial^\mu \Sigma_\Lambda]\\& 
-\frac{c_2}{4\pi}f_\Lambda {\rm Tr}[D_\mu \Sigma_Q^\dagger D^\mu \Sigma_Q]{\rm Tr}[M_\Lambda \Sigma_\Lambda^\dagger +\Sigma_\Lambda M_\Lambda^\dagger -2M_\Lambda E_\Lambda] \\& 
-\frac{c_3}{4\pi}f_\Lambda f {\rm Tr}[M_Q \Sigma_Q^\dagger +\Sigma_Q M_Q^\dagger +2 M_Q E_Q]{\rm Tr}[M_\Lambda \Sigma_\Lambda^\dagger +\Sigma_\Lambda M_\Lambda^\dagger -2M_\Lambda E_\Lambda]\\ &
-\frac{c_4}{4\pi}f {\rm Tr}[M_Q \Sigma_Q^\dagger+\Sigma_Q M_Q^\dagger +2 M_Q E_Q]{\rm Tr}[\partial_\mu \Sigma_\Lambda^\dagger \partial^\mu \Sigma_\Lambda]+\dots\,.
    \end{split}
\end{equation} Here $ c_i $ with $ i=1,\dots,4 $ are the Gasser-Leutwyler type coefficients~\cite{Gasser:1984gg}, where $ c_i\sim \mathcal{O}(1) $ by naive dimensional analysis in analogy with QCD~\cite{Manohar:1983md}. For simplicity, we have shifted the $ c_2 $ and $ c_3 $ terms so that neither $ \Phi $, $ h $, nor the masses and kinetic terms of the EW gauge bosons receive additional contributions from these higher-order terms. Instead, they are determined by Eq.~(\ref{Potential 1}).

To zeroth order depending on vacuum alignment angle $ \theta $, the total potential in Eq.~(\ref{Potential 1}) is given by \begin{equation}
\begin{split}
V_{\rm eff}^0(\theta)=&- C_g f^4 \frac{3g_L^2+g^{2}_Y}{2}c_\theta^2+8\pi f^3  c_Q  m_Q c_\theta  +\frac{C_{tS}}{(4\pi)^2}f^4\Big(y_L^4 s_\theta^4+y_R^4 c_\theta^4+y_L^2y_R^2 s_\theta^2 c_\theta^2\Big) \\ &+\dots\,, \end{split}
\end{equation} where $ m_Q\equiv m_1+m_2 $. By minimizing this potential, $ \partial V_{\rm eff}^0/\partial \theta = 0 $, we can fix the hyper-fermion mass term as a function of the misalignment angle (or vice versa) as follows\begin{equation}
    \label{eq:alignPC}
    m_Q=\frac{f c_\theta}{64\pi^3 c_Q }\bigg(C_{tS}\Big[y_L^4-y_R^4-(y_L^4+y_R^4-y_L^2y_R^2)c_{2\theta}+8\pi^2 C_g(3g_L^2+g_Y^2)\Big]\bigg)\,.
\end{equation} 

By minimizing the total potential in Eq.~(\ref{Potential 1}) with respect to $ \Phi $ and assuming the parametrization in Eq.~(\ref{eq:linear parametrization}), we obtain the VEV of $ \Phi $ (denoted as $ v_\Phi \equiv \sqrt{2}\langle \Phi \rangle $), which, for $ v_\Phi \ll f $, is given by\begin{equation}
    \label{eq:VEV of Phi}
 v_\Phi \approx\frac{8\sqrt{2}\pi c_\Lambda f  \mu}{16\pi c_\Lambda m +C_\Phi f y_\Phi^2} \,.
\end{equation} As anticipated, the VEV tends towards zero when $ \mu $ approaches zero.

%we obtain a condition for the VEV of $ \Phi $ ($  v_\Phi \equiv \langle \Phi \rangle $), which is given by \begin{equation}    \label{eq:alignPC} \mu_\Phi =\frac{16\pi c_\Lambda m +C_\Phi f y_X^2}{8\sqrt{2}\pi c_\Lambda}\frac{v_\Phi}{f}+\mathcal{O}\left[\left(\frac{v_\Phi}{f}\right)^3\right]\,.\end{equation} 

\subsection{The pNGB mass expressions}

For simplicity, we assume that $ m_1 = m_2 = m_Q/2 $ in the following. From the total effective potential in Eq.~(\ref{Potential 1}) and the above vacuum misalignment condition, the physical SM Higgs, $ h\equiv \Pi_Q^4  $, obtains a mass
\begin{align}
    \label{eq: higgs mass}
    m_h^2=\frac{v_{\rm EW}^2}{8\pi^2}\bigg( C_{tS}\Big[y_L^4+3y_R^4-2y_L^2y_R^2+3(y_L^4+y_R^4-y_L^2y_R^2)c_{2\theta}\Big] -8\pi^2 C_g(3g_L^2+g_Y^2) \bigg)\,,
\end{align} and the mass of the pseudo-scalar, $ \eta\equiv \Pi_Q^5 $, is given by \begin{equation}
    \label{eq:}
    m_\eta^2= 8\pi c_Q f m_Q/c_\theta\,.
\end{equation} Moreover, for $ v_\Phi \ll f $, the mass of the pNGB from the breaking of the $ \UU(1)_\Lambda $ symmetry, denoted $ \phi_\Lambda $, achieves the mass\begin{equation}
    \label{eq:}
    m_{\phi_\Lambda}^2\approx 16\pi c_\Lambda f_\Lambda m + \frac{16\sqrt{2}}{3}\pi c_\Lambda v_\Phi \mu \,.
\end{equation}

Finally, the last four pNGBs, $ \Phi_{R,I} $ and $ \Theta_{1,2} $, in Eq.~(\ref{eq:NGBmatrixMod}) mix with each other, resulting in the squared mass matrix in the basis $ ( \phi_{R}, \phi_{I},\Theta_1,\Theta_2) $ for $ v_\Phi \ll f $:  \begin{equation}
    \begin{split} \label{eq: mass matrix Theta eta}
&M_{\Phi\Theta}^2\approx
\begin{pmatrix}
m_{\phi_{R}}^2 & m_{\phi_{R}\phi_{I}}^2 & m_{\phi_{R}\Theta_1}^2 & m_{\phi_{R}\Theta_2}^2 \\
m_{\phi_{R}\phi_{I}}^2 & m_{\phi_{I}}^2  & m_{\phi_{I}\Theta_1}^2  & m_{\phi_{I}\Theta_2}^2
\\
m_{\phi_{R}\Theta_1}^2 & m_{\phi_{I}\Theta_1}^2  & m_{\Theta_1}^2  & m_{\Theta_1\Theta_2}^2
\\
m_{\phi_{R}\Theta_2}^2 & m_{\phi_{I}\Theta_2}^2  & m_{\Theta_1\Theta_2}^2  & m_{\Theta_2}^2
\end{pmatrix}
    \end{split}
\end{equation} with \begin{equation*}
 \begin{split}
&m_{\phi_{R}}^2=-\frac{C_X f^4 y_\Phi^2}{f_\Lambda^2}+16\pi c_\Lambda \left(f_\Lambda m+\sqrt{2}v_\Phi \mu\right) \,, \quad\quad m_{\phi_{R}\phi_{I}}^2 = -\frac{i C_X f^4 y_\Phi^2}{f_\Lambda^2} \\& m_{\phi_{I}}^2=\frac{C_X f^4 y_\Phi^2}{f_\Lambda^2}+16\pi c_\Lambda \left(f_\Lambda m+\sqrt{2}v_\Phi \mu/3\right)\,, \quad\quad \\ &m_{\Theta_1}^2 = \frac{16\pi}{19 f_{\Theta_1}} \left(c_\Lambda f_\Lambda^2 \left(f_\Lambda m+\sqrt{2}v_\Phi \mu\right) +9 c_Q f^3 m_Q c_\theta\right) \,, \quad\quad \\ &m_{\Theta_1\Theta_2}^2= \frac{16\pi c_\Lambda}{\sqrt{19}} \frac{f_\Lambda^2}{f_{\Theta_1}f_{\Theta_2}}\left(f_\Lambda m +\sqrt{2}v_\Phi \mu \right) \,, \quad\quad m_{\Theta_2}^2 = m_\chi^2+16\pi c_\Lambda\frac{f_{\Lambda}^2}{f_{\Theta_2}^2} \left( f_\Lambda m +\sqrt{2}v_\Phi \mu\right)
 \,, \\ & m_{\phi_{R}\Theta_1}^2=2\sqrt{\frac{8}{19}}\frac{f^3 }{f_{\Theta_1}f_\Lambda^2}C_X f v_\Phi y_\Phi^2 \,, \quad\quad m_{\phi_{R}\Theta_2}^2=2\sqrt{2}\frac{f^3 }{f_{\Theta_2}f_\Lambda^2}C_X f v_\Phi y_\Phi^2 \,,\\ & m_{\phi_{I}\Theta_1}^2 = -i m_{\phi_{R}\Theta_1}^2 \,, \quad\quad m_{\phi_{I}\Theta_2}^2 = -i m_{\phi_{R}\Theta_2}^2 \,.
    \end{split}
\end{equation*} If we assume that the explicit mass $ m_\chi $ of the pNGB $ \Theta_2 $ is large, the pNGB $ \Theta_2 $ is decoupled from the other pNGBs. In the following, $ \widetilde{\phi}_{R,I} $ and $ \widetilde{\Theta}_{1,2} $ represent the mass eigenstates of this mass matrix in Eq.~(\ref{eq: mass matrix Theta eta}), predominantly composed of $ \phi_{R,I} $ and $ \Theta_{1,2} $, respectively. 

%%%%%%%%%%%%%%%%%%%%%%%%%%%%%%%%%%%%%%%%%%%%%%%%%%%%%%%%%%%%%%%%%%%%%%%%%%%
\section{Dark matter self-interactions}
\label{sec: Dark matter self-interaction}

The large DM self-interactions are strongly motivated by their ability to solve the small-scale structure problems in our Universe, such as the core-cusp and the "too-big-to-fail" problems. In this model, the DM particles have elastic self-scattering due to the presence of the Yukawa term $ y_\Phi \overline{X} X \Phi $ in Eq.~(\ref{eq: DM mass and Yukawa}). In order to alleviate the small-scale anomalies of $ \Lambda $CDM, the typical DM self-scattering cross-section should be $ \sigma_{\rm SI}\sim 1~\text{cm}^2/\text{g} $, which is fourteen orders of magnitude larger than the typical WIMP cross-section ($ \sigma_{\rm SI}\sim 10^{-14}~\text{cm}^2/\text{g} $). This motivates the existence of a light mediator, which is much lighter than
EW scale. The scalar mediator $ \Phi $ in our model serves this purpose. The
scattering in non-relativistic limit is well-described by the
attractive Yukawa potential, which is given by \begin{equation}
    \label{eq:}
  \frac{y_\Phi^2}{4\pi r} e^{-m_\Phi r}\,.
\end{equation} To encompass the pertinent physics of forward scattering divergence, we introduce the transfer cross-section $ \sigma_T $ as follows~\cite{Feng:2009hw,Tulin:2017ara,Tulin:2013teo} \begin{equation}
    \label{eq:}
  \sigma_T=\int d\Omega (1-\cos \Theta )\frac{d\sigma}{d\Omega}\,.
\end{equation}

\begin{figure}[t]
	\centering
	\includegraphics[width=0.8\textwidth]{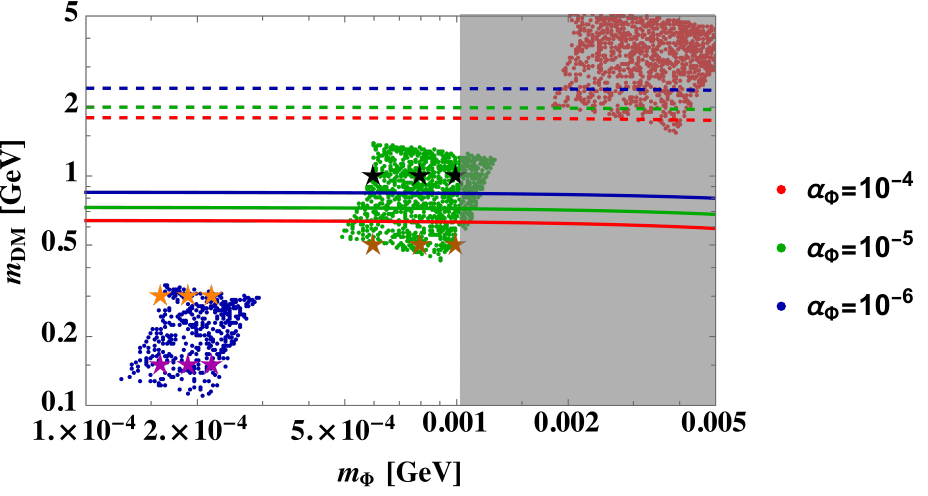}
	\caption{Data points in the $ m_{\Phi}-m_{\rm DM} $ parameter space that result in self-interaction cross sections lying within at least one of the error bars of the measurements from each of the three groups of astrophysical sources: dwarf galaxies, LSBs and galaxy clusters, as obtained from Ref.~\cite{Kaplinghat:2015aga}. The red, green and blue points correspond to $ \alpha_\Phi=10^{-4} $, $ 10^{-5} $ and $ 10^{-6} $, respectively. The stars represent the curves in Figure~\ref{fig: SIDM cross sections} with same color, illustrating the velocity-weighted self-interaction cross sections per unit DM mass as a function of the mean DM collision velocity that align with the observed data. The gray shaded region shows the constraint from BBN. The red, green and blue lines represent the upper constraints on the DM mass for $ \alpha_\Phi =10^{-4}, $ $10^{-5}, $ and $ 10^{-6}, $ respectively, if the value of $ T_{\text{DM}}/T_{\text{VM}} $ set by the relic density constraint is not affected by the portal coupling $ \lambda_{H\Phi} $. The solid and dashed lines denote these constraints corresponding to the coefficients $ c_3=1.0$ and $ 0.1 $ in Eq.~(\ref{eq: double trace terms}), respectively. }
	\label{fig: SIDM cross section parameter space}
\end{figure}

Based on a combination of the masses of the DM ($ m_{\rm DM} $) and the mediator ($ m_{\Phi} $), along with the relative velocity of the colliding particles $ v $ and the coupling ($ \alpha_\Phi\equiv y_\Phi^2/(4\pi) $), three distinct regimes can be identified. In the Born regime ($ \alpha_\Phi m_{\rm DM}/m_\Phi \ll 1 $), perturbative calculations are applicable. In this limit, the transfer cross-section can be written as  \begin{equation}
    \label{eq:}
  \sigma_T^{\rm Born}=\frac{8\pi \alpha_\Phi^2}{m_{\rm DM}^2 v^4}\left[\ln\left(1+\frac{m_{\rm DM}^2v^2}{m_\Phi^2}\right)-\frac{m_{\rm DM}^2v^2}{m_\Phi^2+m_{\rm DM}^2v^2}\right]\,.
\end{equation} Outside the Born regime ($ \alpha_\Phi m_{\rm DM}/m_\Phi \gtrsim 1 $), non-perturbative effects become important and there are two distinct regions, the classical regime and the resonance regime. In the classical regime ($ m_{\rm DM} v /m_\Phi \gg 1 $), the solutions for an attractive potential is given by  \begin{equation}
    \label{eq:}
  \sigma_T^{\rm classical}= \left\{\begin{array}{cc}\frac{4\pi}{m_\Phi^2}\beta^2 \ln\left(1+\beta^{-1}\right), \quad & \beta\lesssim 10^{-1} \\  
  \frac{8\pi}{m_\Phi^2}\beta^2 / \left(1+1.5\beta^{1.65}\right), \quad & 10^{-1}\lesssim\beta\lesssim 10^{3}\\  
  \frac{\pi}{m_\Phi^2} \left(\ln \beta +1-0.5\ln^{-1}\beta\right)^2, \quad & \beta\gtrsim 10^{3}
  \end{array}\right. \,,
\end{equation} where $ \beta\equiv 2\alpha_\Phi m_\Phi /(m_{\rm DM}v^2) $. In the resonant regime ($ m_{\rm DM}v/m_\Phi \lesssim 1 $), the quantum mechanical resonances and anti-resonance in $ \sigma_T $ appear due
to (quasi-)bound states formation in the attractive potential. Within this regime, there exists no analytic formula for $ \sigma_T $, and it must be computed by solving the Schr$ \ddot{\text{o}} $dinger equation directly using a partial wave analysis. However, we can also use the non-perturbative result for $ \sigma_T $ for the Hulth{\'e}n potential, given by $ V(r)=\pm \alpha_\Phi \delta e^{-\delta r}/(1-e^{-\delta r}) $ with the Hulth{\'e}n screening
mass $ \delta = \kappa m_\Phi $ where $ \kappa \approx 1.6 $. The result is  \begin{equation}
    \label{eq:}
  \sigma_T^{\rm Hulth\text{\'e}n}=\frac{16\pi}{m_{\rm DM}^2v^2}\sin^2 \delta_0 \,,
\end{equation} where the $ l=0 $ phase shift $ \delta_0 $ is given in terms of the $ \Gamma $-function by \begin{equation}
    \label{eq:}
  \delta_0 ={\rm arg}\left(\frac{i\Gamma\left(\frac{im_{\rm DM}v}{\kappa m_\Phi}\right)}{\Gamma(\lambda_+)\Gamma(\lambda_-)}\right), \quad \quad \lambda_\pm \equiv 1+\frac{im_{\rm DM}v}{2\kappa m_\Phi}\pm \sqrt{\frac{\alpha_\Phi m_{\rm DM}}{\kappa m_\Phi}-\frac{m_{\rm DM}^2 v^2 }{4\kappa^2 m_\Phi^2}}\,.
\end{equation}

\begin{figure}[t]
	\centering
	\includegraphics[width=0.9\textwidth]{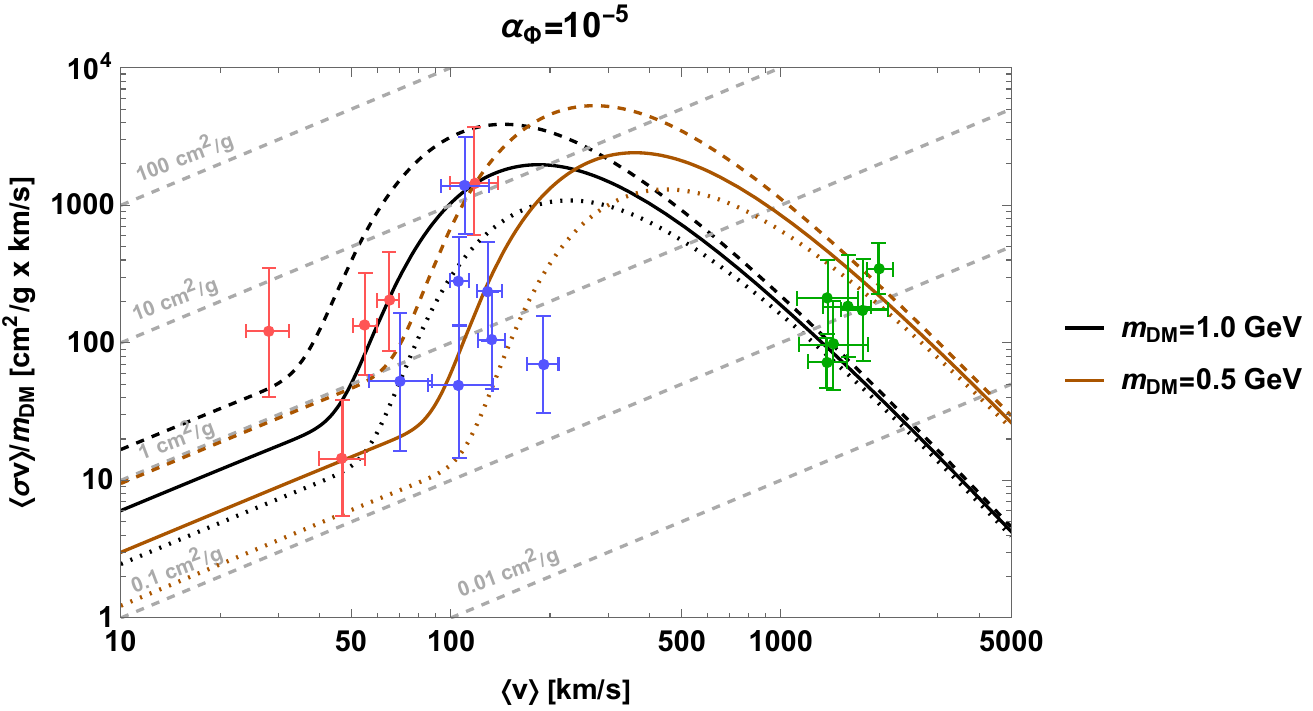} \\
	\quad\\
	\includegraphics[width=0.9\textwidth]{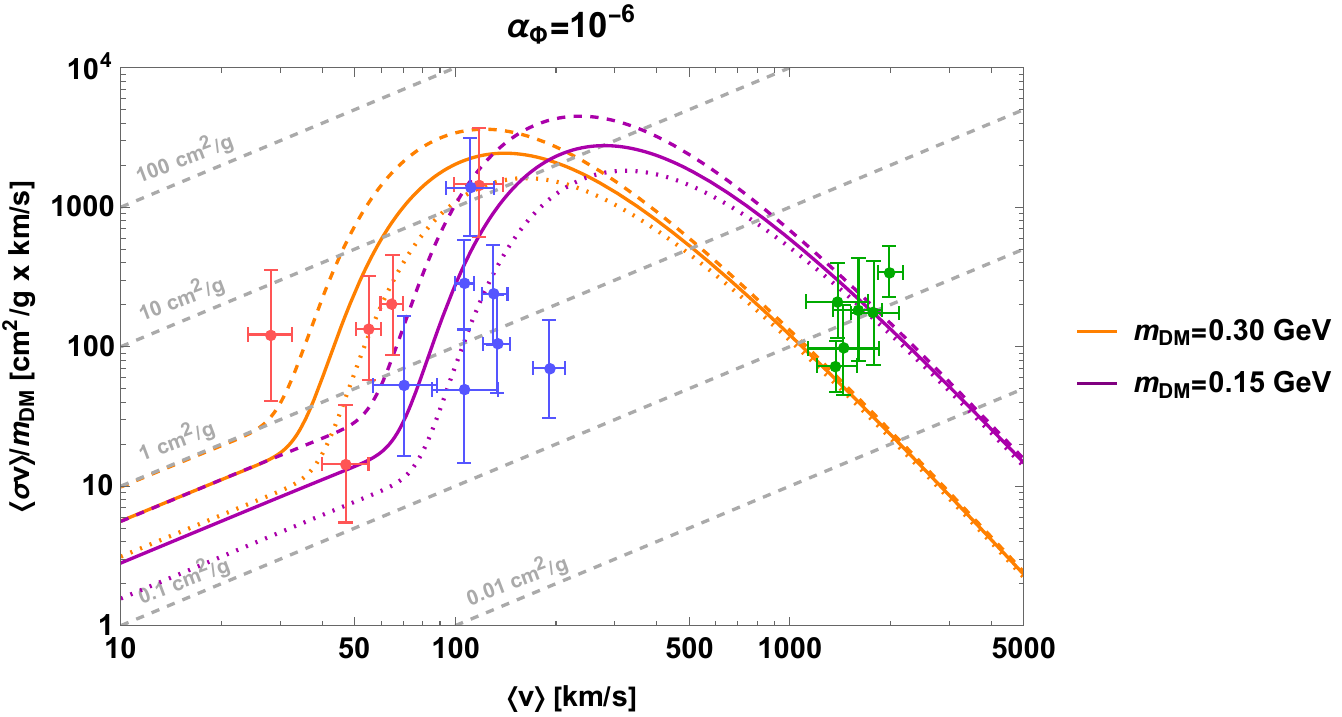}
	\caption{The velocity-weighted self-interaction cross sections per unit DM mass as a function of mean DM collision velocity for $ \alpha_\Phi=10^{-5} $ (upper) and $ \alpha_\Phi=10^{-6} $ (lower panel). The data points, obtained from Ref.~\cite{Kaplinghat:2015aga}, correspond to measurements from various astrophysical sources: dwarf galaxies (red), LSBs (blue) and galaxy clusters (green). The diagonal lines represent contours of constant $ \sigma/m_{\rm DM} $ (cross section per unit mass). For the various DM masses ($m_{\rm DM} = 1.0,0.5 $~GeV for $ \alpha_\Phi=10^{-5} $ and $ 0.30,0.15 $~GeV for $ \alpha_\Phi=10^{-6} $), the dotted, solid and dashed lines depict the predictions from the model, considering the various mediator masses ($ m_\Phi =1.00 $, $ 0.75 $ and $ 0.50 $~MeV, respectively.  }
	\label{fig: SIDM cross sections}
\end{figure}

In Figure~\ref{fig: SIDM cross section parameter space}, we present data points in the parameter space of $ m_{\Phi}-m_{\rm DM} $, where self-interaction cross sections fall within at least one of the error bars of the measurements from three distinct groups of astrophysical sources: dwarf galaxies, LSBs, and galaxy clusters, as obtained from Ref.~\cite{Kaplinghat:2015aga}. Here, the red, green and blue points correspond to the couplings $ \alpha_\Phi=10^{-4} $, $ 10^{-5} $ and $ 10^{-6} $, respectively. The gray shaded region shows the constraint from BBN~\cite{Hambye:2019tjt}, while all parameter space is allowed by CMB, indirect and direct detection experiments. The red, green and blue lines represent the upper bounds on the DM mass for $ \alpha_\Phi =10^{-4}, $ $10^{-5}, $ and $ 10^{-6}, $ respectively, under the assumption that the relic density constraint is not influenced by the portal coupling $ \lambda_{H\Phi} $, which will be discussed in the next section. The solid and dashed lines denote these constraints corresponding to the coefficients $ c_3=1.0$ and $ 0.1 $ in Eq.~(\ref{eq: double trace terms}), respectively. Finally, the stars represent the curves in Figure~\ref{fig: SIDM cross sections} with same color, illustrating the velocity-weighted self-interaction cross sections per unit DM mass as a function of the mean DM collision velocity that align with the observed data. 

%%%%%%%%%%%%%%%%%%%%%%%%%%%%%%%%%%%%%%%%%%%%%%%%%%%%%%%%%%%%%%%%%%%%%%%%%%%
\section{Dark matter relic density and various constraints}
\label{sec: Dark matter relic density and various constraints}

\begin{figure}[t]
	\centering
	\includegraphics[width=0.95\textwidth]{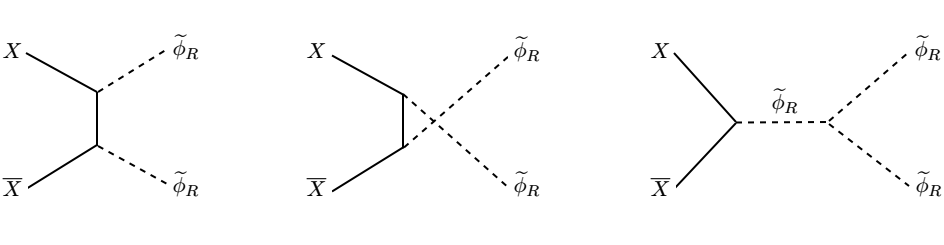}
	\caption{The Feynman diagrams represent the various annihilation processes, $ X\overline{X}\leftrightarrow \widetilde{\phi}_R \widetilde{\phi}_R $, which determine the relic density of DM. }
	\label{fig: Feynman Diagrams}
\end{figure}

The primary constraints on hidden particles with masses $ \lesssim \mathcal{O}(\text{MeV}) $ (e.g. the mediator particle in this model) arise from measurements of the relative abundance of light elements from BBN~\cite{Aver:2015iza,Peimbert:2016bdg}, as well as measurements of the CMB power spectrum~\cite{Planck:2015fie,Planck:2018vyg}. From these constraints, the model is only marginally allowed, with a narrow region particularly around $ m_{\rm DM}\sim 0.5 $~GeV and $ m_{\Phi}\sim 1.1 $~MeV remaining unexcluded for this model~\cite{Hufnagel:2018bjp}. This narrow allowed parameter space can open up as soon as we allow for DM to thermalize within a hidden sector with a temperature $ T_{\rm DM} $ lower than the temperature of the visible sector $ T_{\rm VM} $. This possibility could explain the DM relic density, where an annihilation cross section smaller than the usual thermal freeze-out scenario arises if both the hidden sector light mediator and DM particles thermalize (forming a thermal bath with temperature $ T_{\rm DM} $) but do not with the visible sector with temperature $ T_{\rm VM} $, resulting in $ T_{\rm DM}/T_{\rm VM} < 1 $. 

The largest value of relic number density for DM particles achievable today cannot surpass the quantity that existed when these particles were relativistic. The value is saturated if dark matter decouples relativistically. In such instances, the number density of the DM particles at the temperature when they decoupled is $ n_{\rm DM}=\zeta(3)g_{\rm DM}^{\rm eff} T_{\rm DM}^3/\pi^2 $, and consequently, it is suppressed by $ (T_{\rm DM}/T_{\rm VM})^3 $. As shown in Ref.~\cite{Hambye:2019tjt}, requiring that it yields a relic density equal to or greater than the observed value, we obtain the lower bound\begin{equation}
    \label{eq: bound 1}
\frac{T_{\rm DM}}{T_{\rm VM}}\geq 1.14\times 10^{-3} \left(\frac{1~\text{GeV}}{m_{\rm DM}}\right)^{1/3}\left(\frac{g_*^S(T_{\rm VM,dec})}{g_{\rm DM}^{\rm eff}(T_{\rm DM,dec})}\right)^{1/3}\,,
\end{equation} where $ T_{\rm DM,dec} $ and $ T_{\rm VM,dec} $ represent the values of $ T_{\rm DM} $ and $ T_{\rm VM} $, respectively, at the time when DM decoupled. Furthermore, a bound on DM interactions, $ \alpha_\Phi $, is valuable when necessitating that the DM sector thermalizes before DM decoupled, enabling the definition of a DM sector temperature denoted as $ T_{\rm DM} $. This constraint can be approximated by ensuring that the inequality $ \Gamma/H $ at $ T_{\rm DM}=m_{\rm DM} $ is satisfied, which results in \begin{equation}
    \label{eq: bound 2}
\alpha_\Phi\geq 3.95\times 10^{-10} \frac{T_{\rm VM}}{T_{\rm DM}} \left(\frac{\sqrt{g_{\rm DM}^{\rm eff}(T_{\rm VM})}}{g_{\rm DM}^{\rm eff}(T_{\rm DM})}\right)^{1/2}\left(\frac{m_{\rm DM}}{1~\text{GeV}}\right)^{1/2}\,.
\end{equation} 

In the following, we compute the thermal relic density of the DM. Based on the Feynman diagrams shown in Figure~\ref{fig: Feynman Diagrams}, the thermal-averaged annihilation cross-section of the DM is given by \begin{equation}
\begin{split} 
    \label{eq: annihilation cross-section}
\langle\sigma v\rangle = & \frac{m_{\rm DM}y_\Phi^2 \langle v^2 \rangle}{96\pi (m_\Phi^2-2m_{\rm DM}^2)^4(m_{\rm DM}m_\Phi^2-4m_{\rm DM}^3)^2}\sqrt{m_{\rm DM}^2-m_\Phi^2}\Big[3 \lambda_{3\Phi}^2 (m_\Phi^2-2m_{\rm DM}^2)^4 \\ &+4 y_\Phi\lambda_{3\Phi}m_{\rm DM}(28 m_{\rm DM}^4-11m_{\rm DM}^2 m_\Phi^2+m_\Phi^4 )(m_\Phi^2-2m_{\rm DM}^2)^2 \\& +8 g_\Phi^2 m_{\rm DM}^2 (m_\Phi^2-4m_{\rm DM}^2)^2(9m_{\rm DM}^4 -4 m_{\rm DM}^2 m_\Phi^2 +m_\Phi^4)\Big]\,,
\end{split} 
\end{equation} where $ \lambda_{3\Phi} $ is the coupling of the cubic term for the mediator (identified as the mass eigenstate $ \widetilde{\phi}_R $) in the total effective potential in Eq.~(\ref{Potential 1}). Therefore, the annihilation is dominated by p-wave process. As found in Ref.~\cite{Chacko:2015noa}, the freeze-out temperature can be calculated from \begin{equation}
\begin{split} 
    \label{eq: freeze-out temperature}
x_f = \frac{T_{\rm DM}}{T_{\rm VM}}\log\left[\frac{9}{\pi^3}\sqrt{\frac{5}{2}}\frac{g_{\rm DM}}{\sqrt{g_*^{\rm eff}}}m_{\rm DM}m_{\rm P} \left(\frac{T_{\rm DM}}{T_{\rm VM}}\right)^{7/2}\frac{\langle\sigma v\rangle}{\langle v^2\rangle }\frac{1}{x_f^{3/2}}\right]\,, 
\end{split} 
\end{equation} where $ m_P=1.22\times 10^{19} $~GeV is the Planck mass and $ x_f\equiv m_{\rm DM}/T $. Finally, the relic density of DM is expressed as \begin{equation}
\begin{split} 
    \label{eq: relic density of DM}
\Omega_{\rm DM}h^2 = m_{\rm DM}s_0 Y_{\infty}\rho_c^{-1}
\end{split} 
\end{equation} with \begin{equation}
\begin{split} 
    \label{eq:}
Y_\infty =\left(\sqrt{\frac{\pi}{5}}\frac{g_*}{\sqrt{g^{\rm eff}_*}}m_{\rm P}m_{\rm DM}\frac{T_{\rm DM}}{T_{\rm VM}}\frac{\langle\sigma v\rangle}{\langle v^2\rangle}\frac{1}{ x_f^2}\right)^{-1}\,,
\end{split} 
\end{equation} where $ s_0 $ and $ \rho_c $ denote the present-time entropy density and critical density, respectively.

When $T_{\rm DM}/T_{\rm VM}<1$ at freeze-out, there is a minor additional effect stemming from the dependence of the argument of the logarithm in Eq.~(\ref{eq: freeze-out temperature}) on $T_{\rm DM}/T_{\rm VM}$. However, this effect is small. Therefore, the value of $ \langle\sigma v\rangle / \langle v^2\rangle $ in Eq.~(\ref{eq: annihilation cross-section}) that corresponds to the observed $ \Omega_{\rm DM} $ is reduced by a factor of $ T_{\rm DM}/T_{\rm VM} $ compared to the case of the standard thermal cross section, which is approximately $3\times 10^{-26}~\text{cm}^3/\text{s}$. 
 
\begin{figure}[t]
	\centering
	\includegraphics[width=0.7\textwidth]{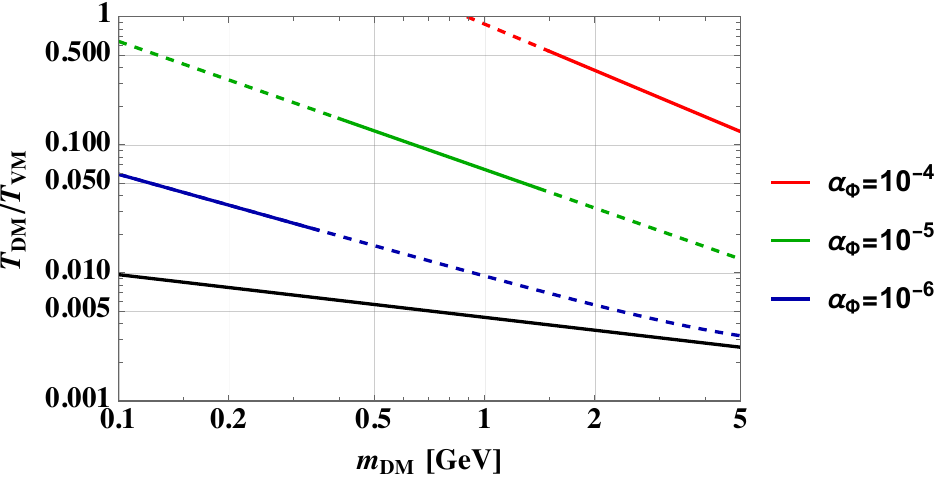}
	\caption{Values of $ T_{\rm DM}/T_{\rm VM} $ needed in order to reproduce the observed relic abundance of DM as a function of the DM mass for couplings $ \alpha_\Phi=10^{-4} $ (red), $ 10^{-5} $ (green) and $ 10^{-6} $ (blue line). The dashed lines fall outside the range of the DM mass displayed in Figure~\ref{fig: SIDM cross section parameter space}, resulting in self-interactions that are not consistent with the observed data.  }
	\label{fig: relic density DM temperature}
\end{figure} 
 
Figure~\ref{fig: relic density DM temperature} depicts the necessary values of $ T_{\rm DM}/T_{\rm VM} $ for reproducing the observed relic abundance of DM from Eq.~(\ref{eq: relic density of DM}) across different DM masses. The colored lines represent various couplings: $ \alpha_\Phi=10^{-4} $ (red), $ \alpha_\Phi=10^{-5} $ (green), and $ 10^{-6} $ (blue). Here, the solid lines intersect with the range of DM mass shown in Figure~\ref{fig: SIDM cross section parameter space}, leading to self-interactions that align with the observed data within a range of mediator mass values, approximately $ 2~\text{MeV}<m_\Phi < 6~\text{MeV} $ for $ \alpha_\Phi=10^{-4} $, $ 0.5~\text{MeV}<m_\Phi < 1.5~\text{MeV} $ for $ \alpha_\Phi=10^{-5} $ and $ 0.1~\text{MeV}<m_\Phi < 0.3~\text{MeV} $ for $ \alpha_\Phi=10^{-6} $. In the calculations of the DM relic density, these small mediator mass values have negligible effects on the values of $ T_{\rm DM}/T_{\rm VM} $ in the figure. Furthermore, the black line in Figure~\ref{fig: relic density DM temperature} represents the lower bound on $ T_{\rm DM}/T_{\rm VM} $ given in Eq.~(\ref{eq: bound 1}), which is satisfied in the figure. The additional constraint in Eq.~(\ref{eq: bound 2}), resulting in an upper bound on the DM mass, lies outside the plotted range in the figure (with $ m_{\rm DM}\lesssim 2\times 10^3 $~GeV for $ \alpha_\Phi =10^{-4} $, $ m_{\rm DM}\lesssim 10^2 $~GeV for $ \alpha_\Phi =10^{-5} $, and $ m_{\rm DM}\lesssim 8 $~GeV for $ \alpha_\Phi =10^{-6} $) and is also satisfied. 

Finally, the results presented in Figure~\ref{fig: relic density DM temperature} are derived under the assumption of no portal interaction. If the portal interaction is activated, these results remain largely unchanged as long as the portal is sufficiently small to avoid heating the hidden sector beyond the temperature $ T_{\rm DM} $ assumed above. In Figure~\ref{fig: SIDM cross section parameter space}, the red, green, and blue lines depict the upper bounds on the DM mass for $ \alpha_\Phi =10^{-4}, $ $10^{-5}, $ and $ 10^{-6}, $ respectively. These bounds are imposed to ensure that the portal interaction never results in numbers of DM and light mediator particles exceeding those obtained without portal coupling. The solid and dashed lines represent these constraints for the coefficients $ c_3=1.0$ and $ 0.1 $ in Eq.~(\ref{eq: double trace terms}), respectively. Although there are regions in parameter space where the portal coupling exceeds these values (particularly along reannihilation and secluded freeze-out pathways, where $ T_{\rm DM} $ is primarily influenced by the energy transferred through the portal~\cite{Chu:2011be,Hambye:2019dwd}), for simplicity, we have calculated the DM relic density assuming that the portal coupling has a negligible effect on it.

As follows, we address the additional constraints imposed by CMB, BBN, as well as direct and indirect detection. First, the p-wave nature of DM annihilation in this model avoids the CMB constraint on the DM annihilation cross section. In this scenario, the annihilation is significantly less boosted at recombination time compared to the s-wave case and remains consistent with the CMB constraint, as illustrated in Figure~1 of Ref.~\cite{Bringmann:2016din}. As discussed in Ref.~\cite{Hambye:2019tjt}, the CMB constraints on the decay of the light mediator are also alleviated when $ T_{\rm DM}/T_{\rm VM} < 1$, as the number of remaining mediators upon decoupling from the DM particles is reduced by a factor of $ (T_{\rm DM}/T_{\rm VM})^3 $ compared to the case where $ T_{\rm DM}/T_{\rm VM} =1 $. Furthermore, when $ T_{\rm DM}/T_{\rm VM} <1 $, the CMB constraint on $ N_{\text{eff}} $ becomes quickly irrelevant, as the number density of light mediators is reduced by a factor of $ (T_{\rm DM}/T_{\rm VM})^3 $, see Table~II in Ref.~\cite{Hambye:2019tjt}. Secondly, the constraints from BBN, including the Hubble constant/entropy injection and photodisintegration constraints, are also alleviated due to the $ (T_{\rm DM}/T_{\rm VM})^3 $ suppression of the light mediator number density. However, for $ m_\Phi >2m_e $, the decay width of the light mediator is suppressed by the electron Yukawa coupling. Consequently, the upper bound on the light mediator lifetime from BBN requires a relatively large value of the portal coupling, potentially leading to thermalization of the hidden sector, which is in tension with $ T_{\rm DM}/T_{\rm VM}<1 $. Nevertheless, the parameter space in Figure~\ref{fig: SIDM cross section parameter space} is now considerably open for $ m_\Phi <2m_e $, while the gray shaded region shows this constraint in the figure. Moreover, the constraint from the non-overclosure of the light mediator imposes an upper bound on $ T_{\rm DM}/T_{\rm VM} $, which lies above all the lines shown in Figure~\ref{fig: relic density DM temperature}. This is due to the fact that $ m_\Phi \ll m_{\rm DM} $ (see Eq.~(12) in Ref.~\cite{Hambye:2019tjt}). Finally, both direct and indirect detection signals are diminished due to the fact that the coupling between DM and the mediator is even smaller than needed in the case where $ T_{\rm DM}/T_{\rm VM}=1 $.

%In Figure~\ref{fig: SIDM cross section parameter space}, the red, green and blue lines represent the upper bounds on the DM mass for $ \alpha_\Phi =10^{-4}, $ $10^{-5}, $ and $ 10^{-6}, $ respectively, imposing that the portal interaction never leads to numbers of DM and light mediator particles which are larger than the ones one has without portal coupling. The solid and dashed lines denote these constraints corresponding to the coefficients $ c_3=1.0$ and $ 0.1 $ in Eq.~(\ref{eq: double trace terms}), respectively.

%%%%%%%%%%%%%%%%%%%%%%%%%%%%%%%%%%%%%%%%%%%%%%%%%%%%%%%%%%%%%%%%%%%%%%%%%%%
\section{Numerical results of the minimal model}
\label{sec: Model results}

First, we demonstrate that the composite models introduced in this paper can develop two different dynamical scales, decay constants of the visible and dark matter sectors, denoted as $ f $ and $ f_\Lambda $, respectively, similar to Ref.~\cite{Frandsen:2011kt}. Furthermore, we assume that $ T_{\rm DM}/T_{\rm VM} \sim f_\Lambda/f $ since the two sector are almost decoupled, ensuring $ T_{\rm DM}/T_{\rm VM}<1 $ as wanted.

 \begin{figure}[t]
	\centering
	\includegraphics[width=0.6\textwidth]{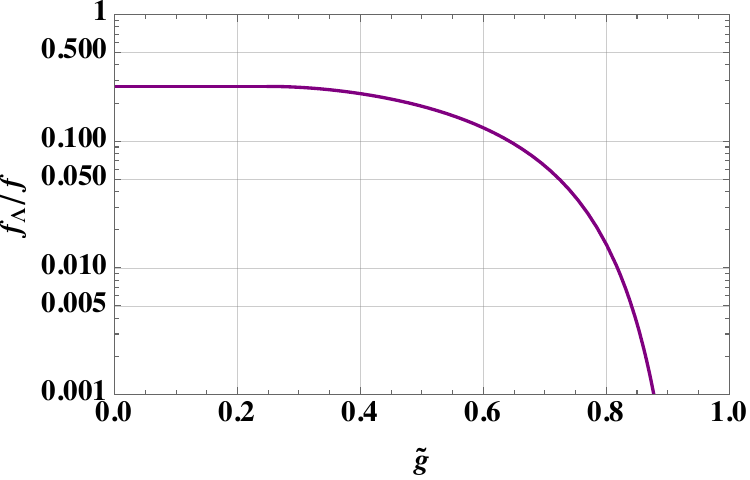}
	\caption{The scale ratio between $ f_\Lambda $ and $ f $ as a function of the coupling $ \widetilde{g} $ for $ G_{\rm HC}=\SP(4) $. }
	\label{fig: scale ratio}
\end{figure}

In Ref.~\cite{Frandsen:2011kt}, they adopted the gauged Nambu-Jona-Lasanio (NJL) model~\cite{Yamawaki:1996vr} as a representative simple theory with four-fermion interactions which leads to chiral symmetry breaking. In our case, the gauged NJL model for the hyper-fermions $ Q $ transforming under fundamental representation of the hypercolor gauge group $ G_{\rm HC}=\SP(2N) $ can be written by extending the composite Higgs Lagrangian $ \mathcal{L}_{\rm CH} $ in Eq.~(\ref{eq: Basic Lagrangian (UV)}) as follows \beq 
\mathcal{L}_{\rm CH} \rightarrow \mathcal{L}_{\rm CH}+\frac{\pi^2 \widetilde{g}}{2 \widetilde{\Lambda}^2 }\left[(\overline{Q}Q)^2+(\overline{Q}i\gamma_5 T^a Q)^2\right] \,, 
\label{eq:}
\eeq where $ \widetilde{g} $ represents a dimensionless four-fermion coupling and $ \widetilde{\Lambda} $ denotes the effective cut-off of the four-fermion operator. The critical value of the coupling for chiral symmetry breaking, as a function of $ \widetilde{g} $, is~\cite{Kondo:1988qd} \beq 
\alpha_{\rm HC}^c (\mathcal{R}_1,\widetilde{g})=\left\{ \begin{array}{cl}
4\left(\sqrt{\widetilde{g}}-\widetilde{g}\right)\alpha_{\rm HC}^c(\mathcal{R}_1) \ & \text{ for } 1/4 < \widetilde{g} \leq 1\,, \\
\alpha_{\rm HC}^c(\mathcal{R}_1) \ & \text{ for } 0 \leq \widetilde{g} < 1/4\,, 
\end{array}\right.
\label{eq:}
\eeq where we use the ladder approximation to the Schwinger-Dyson equation, giving rise to $ \alpha_{\rm HC}^c(\mathcal{R}) = \pi /(3C_2(\mathcal{R})) $ with the quadratic Casimir $ C_2(F)=(N-1)/4$ and $ C_2(A)=N-1$ for the fundamental and adjoint representations of $ G_{\rm HC}=\SP(2N) $ for the visible and DM sectors, respectively. Integrating
the one-loop beta-function $ \beta(\alpha_{\rm HC})=-\left(\tfrac{\alpha^2}{2\pi}\beta_0+\tfrac{\alpha^3}{8\pi^2}\beta_1+\dots\right) $ from $ f $ to $ f_\Lambda $ gives rise to the ratio of the scales, expressed as: \beq 
\frac{f}{f_\Lambda}\simeq \exp\left[\frac{2\pi}{\beta_0(\mathcal{R}_2)}\bigg(\alpha_{\rm HC}^c(\mathcal{R}_1)^{-1}-\alpha_{\rm HC}^c(\mathcal{R}_2)^{-1}\bigg)\right]\,.
\label{eq:}
\eeq Since $ C_2(A)>C_2(F) $, $ f>f_\Lambda $ as desired. 

%Moreover, the estimate of $ \alpha_{\rm HC}^c $ should be compared to the two-loop fixed point value of the coupling $ \alpha_*=-4\pi \beta_0/\beta_1 $. If $ \alpha_{\rm HC}^* <\alpha_{\rm HC}^c $, the theory will run to an infra-red fixed point before triggering chiral symmetry breaking. The lower boundary of the conformal window is thus identified by demanding $ \alpha_{\rm HC}^*(\mathcal{R}_1,\mathcal{R}_2) =\alpha_{\rm HC}^c(\mathcal{R}_1) $. 

In Figure~\ref{fig: scale ratio}, the ratio $ f_\Lambda/f \sim T_{\rm DM}/T_{\rm VM} $ is shown for varying coupling $ \widetilde{g} $ in the scenario with $ G_{\rm HC}=\SP(4) $. Thus, the model appears to dynamically induce the required ratio of temperatures for the two sectors, as depicted in Figure~\ref{fig: relic density DM temperature}, for all the parameter space corresponding to the viable values $ \alpha_\Phi =10^{-5} $ and $ 10^{-6} $ in Figure~\ref{fig: SIDM cross section parameter space}.

Finally, we calculate the masses of the composite pNGBs and the hyper-fermion masses. In our calculations, we assume specific values for the coefficients in the effective potential (see Eq.~(\ref{Potential 1})): $c_Q$ and $c_\Lambda$ are both set to $ 1.5 $, a choice supported by lattice simulations~\cite{Arthur:2016dir}; $C_g$ and $C_X$ are set to $ 1 $; and $C_{tS}$ in Eq.~(\ref{eq: LO terms}) is determined by the Higgs mass, set to approximately $ 0.1 $. For the operators in Eq.~(\ref{eq: top Yukawa operator}) responsible for generating the top mass and Yukawa coupling, we assume $C_{yS} = 1$ and $y_L = y_R$ for simplicity. Additionally, we choose $s_\theta = 0.1$ in our calculations.

%In Figure~\ref{fig: pNGB Masses}, the pNGB masses of $ \widetilde{\phi}_I $ (left) and $ \phi_\Lambda $ (right panel) for varying DM mass with couplings $ \alpha_\Phi=10^{-4} $ (red), $ \alpha_\Phi=10^{-5} $ (green) and $ 10^{-6} $ (blue line). We assume that $T_{\rm DM}/T_{\rm VM}= f_\Lambda /f $ with the values of $ T_{\rm DM}/T_{\rm VM} $ from Figure~\ref{fig: relic density DM temperature}, resulting in the observed relic abundance of DM. The masses of the other pNGBs are (almost) independent of the DM mass and coupling, maintaining values of $ m_{\widetilde{\Theta}_1}\simeq 261 $~GeV, $ m_{\eta}= 270 $~GeV, with $ m_h $ fixed at $ 125 $~GeV. All these masses are almost independent of the mediator mass, denoted as $ m_\Phi \equiv m_{\widetilde{\phi}_R} $. The mass eigenstates $ \widetilde{\phi}_{R,I} $ and $ \widetilde{\Theta}_{1} $ consist mostly of $ \phi_{R,I} $ and $ \Theta_{1} $, respectively. The dashed lines fall inside the range of the DM mass displayed in Figure~\ref{fig: SIDM cross section parameter space}, leading to self-interactions that are consistent with the observed data.

\begin{figure}[t]
	\centering
	\includegraphics[width=0.443\textwidth]{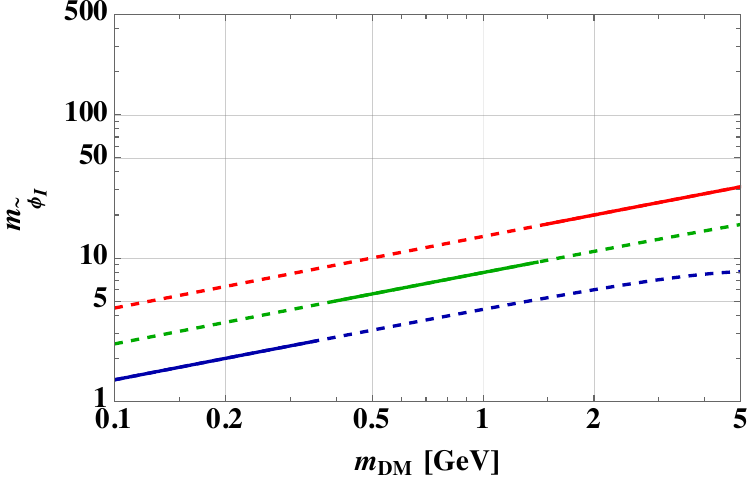}
	\includegraphics[width=0.548\textwidth]{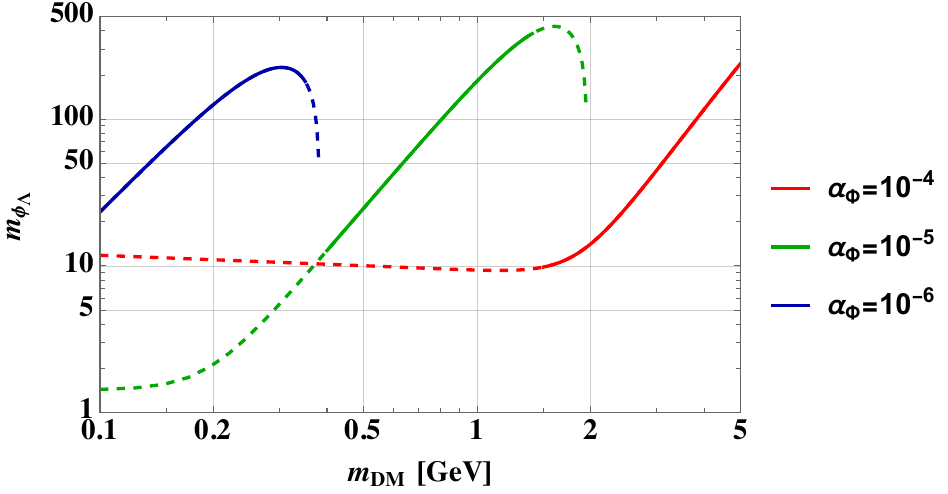}
	\caption{Masses of the pNGBs $ \widetilde{\phi}_I $ (left) and $ \phi_\Lambda $ (right panel) as a function of DM mass with couplings $ \alpha_\Phi=10^{-4} $ (red), $ \alpha_\Phi=10^{-5} $ (green) and $ 10^{-6} $ (blue lines), while $ s_\theta=0.1 $. We assume that $T_{\rm DM}/T_{\rm VM}= f_\Lambda /f $ with the values of $ T_{\rm DM}/T_{\rm VM} $ from Figure~\ref{fig: relic density DM temperature}, resulting in the observed relic abundance of DM. The masses of the other pNGBs are (almost) independent of the DM mass and coupling, maintaining values of $ m_{\widetilde{\Theta}_1}\simeq 261 $~GeV, $ m_{\eta}= 270 $~GeV, with $ m_h $ fixed at $ 125 $~GeV. All these masses are almost independent of the mediator mass, $ m_\Phi \equiv m_{\widetilde{\phi}_R} $. The dashed lines fall within the range of the DM mass displayed in Figure~\ref{fig: SIDM cross section parameter space}, leading to DM self-interactions that are not consistent with the observed data.}
	\label{fig: pNGB Masses}
\end{figure}

In Figure~\ref{fig: pNGB Masses}, the pNGB masses of $ \widetilde{\phi}_I $ (left panel) and $ \phi_\Lambda $ (right panel) vary with the DM mass, considering couplings $ \alpha_\Phi=10^{-4} $ (red), $ \alpha_\Phi=10^{-5} $ (green) and $ \alpha_\Phi=10^{-6} $ (blue lines). We assume a ratio of the temperatures in the visible and dark matter sectors is given by $T_{\rm DM}/T_{\rm VM} = f_\Lambda / f $, with the specific values of $ T_{\rm DM}/T_{\rm VM} $ obtained from Figure~\ref{fig: relic density DM temperature}, ensuring the resulting relic abundance of DM matches observations. It is noted that the masses of the additional pNGBs, namely $ \widetilde{\Theta}_1 $ and $ \eta $, remain (almost) independent of the DM mass and coupling, maintaining values of $ m_{\widetilde{\Theta}_1} \simeq 261 $ GeV and $ m_{\eta} = 270 $ GeV, with $ m_h $ fixed at $ 125 $ GeV. The mass eigenstates, $ \widetilde{\phi}_{R,I} $ and $ \widetilde{\Theta}_{1} $, predominantly consist of $ \phi_{R,I} $ and $ \Theta_{1} $, respectively. All these masses exhibit almost no dependence on the mediator mass, $ m_\Phi \equiv m_{\widetilde{\phi}_R} $ and therefore we have ignored such a dependence. Furthermore, the solid lines in the figure fall within the range of DM mass displayed in Figure~\ref{fig: SIDM cross section parameter space}, leading to DM self-interactions that align with the observed data.

\begin{figure}[t]
	\centering
	\includegraphics[width=0.7\textwidth]{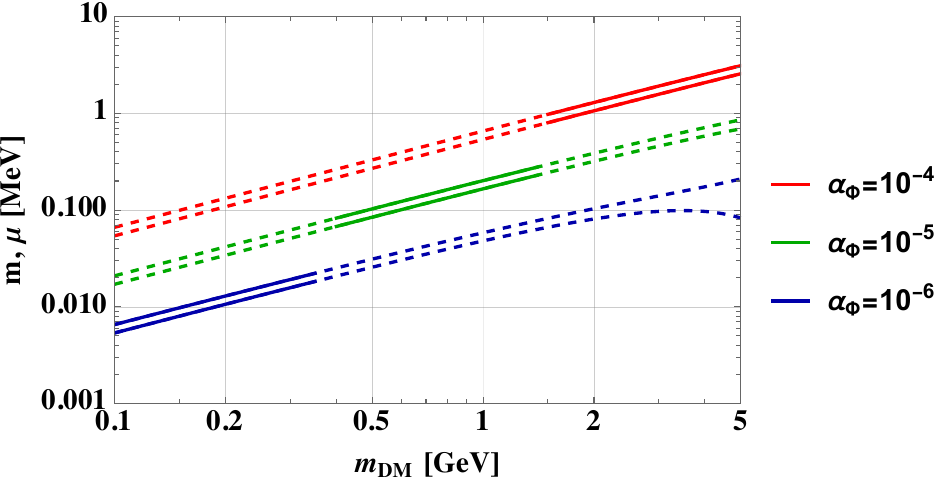}
	\caption{The gauge-invariant hyper-fermion masses, $ \mu $ (upper) and $ m $ (lower lines), of the hidden sector in Eq.~(\ref{eq: hyper-fermion masses}) for varying DM mass with couplings $ \alpha_\Phi=10^{-4} $ (red), $ \alpha_\Phi=10^{-5} $ (green) and $ 10^{-6} $ (blue lines), while $ s_\theta=0.1 $. We assume that $T_{\rm DM}/T_{\rm VM}= f_\Lambda /f $ with the values of $ T_{\rm DM}/T_{\rm VM} $ from Figure~\ref{fig: relic density DM temperature}, resulting in the observed relic abundance of DM. The hyper-fermion masses, $ m_{1,2} $, in Eq.~(\ref{eq: hyper-fermion masses}) of the visible sector are independent of the DM mass and coupling, with the value $ m_Q = m_{1,2}/2 = 0.78 $~GeV. }
	\label{fig: Vector-Like Masses}
\end{figure}

In Figure~\ref{fig: Vector-Like Masses}, the hyper-fermion masses are shown as a function of the DM mass in similar way as the pNGB masses in Figure~\ref{fig: pNGB Masses}. For the lines with same color, the upper and lower lines represent hyper-fermion masses $ \mu $ and $ m $, respectively. Furthermore, the hyper-fermion masses, $ m_{1,2} $, in Eq.~(\ref{eq: hyper-fermion masses}) of the visible sector are independent of the DM mass and coupling, with the value $ m_Q = m_{1,2}/2 = 0.78 $~GeV.  Again all these masses are almost independent of the mediator mass, $ m_\Phi \equiv m_{\widetilde{\phi}_R} $. The masses $\mu$ and $m$ are determined through the minimization of the potential in the direction of the mediator and fixing its mass $m_\Phi$. Meanwhile, the hyper-fermion mass $m_Q$ is obtained in Eq.~(\ref{eq:alignPC}) through minimization in the Higgs direction. Notably, as indicated by Eq.~(\ref{eq: mass matrix Theta eta}), achieving a low mass for the light mediator necessitates a tuning between the contributions from the DM fermion loop and the hyper-fermion masses, $\mu$ and $m$. Even with the smaller value of $f_\Lambda$ compared to $f$, this tuning is significant. As the mediator, a pNGB, is nearly massless, this limit is technically natural according to 't Hooft's naturalness principle~\cite{tHooft:1979rat}, as it reveals the restoration of a global symmetry. 

%%%%%%%%%%%%%%%%%%%%%%%%%%%%%%%%%%%%%%%%%%%%%%%%%%%%%%%%%%%%%%%%%%%%%%%%%%%
\section{Conclusions}
\label{sec: Conclusions}

In conclusion, our investigation explores a novel framework wherein the mass of a self-interacting dark matter (SIDM) candidate, specifically a Dirac fermion, is generated through composite dynamics within a minimal Composite Higgs (CH) model. In these models, a light scalar mediator emerges alongside the Higgs as composite particles, providing a compelling solution to both the halo structure problems at various scales and the Standard Model (SM) naturalness problem. The relic density of the dark matter (DM) candidate is shown to be particle anti-particle symmetric and results from thermal freeze-out.

We address the challenges posed to the traditional Weakly Interacting Massive Particle (WIMP) paradigm, highlighting the core-cusp, ``too big to fail'' and ``missing satellite'' problems at various scales. Our model incorporates SIDM with a velocity-dependent cross-section, successfully alleviating these issues. The emergence of the light composite mediator allows for elastic scattering among DM particles, leading to a range of cross-section values compatible with both small-scale structures and observations of larger systems.

Crucially, our study demonstrates that the Composite Higgs framework provides a natural setting for the emergence of both the Higgs and the dark mediator particles. The CH model allows for the almost decoupled evolution of the two sectors, with relevant global symmetries arising from a confining four-dimensional gauge-fermion theory. This common composite dynamics origin sheds light on the connection between the Higgs and SIDM, addressing shortcomings in the SM description of visible mass and offering a potential resolution to the naturalness and triviality problems associated with the elementary Higgs particle.

Finally, we explore the viable parameter space of our concrete model, ensuring consistency with various constraints set by DM relic density, Big Bang Nucleosynthesis, Cosmic Microwave Background, as well as direct and indirect detection experiments. All these constraints can be mitigated by having the temperature of the DM sector smaller than that of the visible sector ($ T_{\rm DM}/T_{\rm VM} <1$). This naturally occurs because the composite models introduced in this paper can develop two different dynamical scales, $ f_\Lambda/f < 1 $, while the two sectors are almost decoupled, ensuring $ T_{\rm DM}/T_{\rm VM}<1 $ as wanted. In conclusion, the results of our numerical analysis in this paper support the feasibility of our minimal CH model with SIDM, providing a promising avenue for future research in the quest to unravel the mysteries of DM and its interactions. 

%%%%%%%%%%%%%%%%%%%%%%%%%%%%%%%%%%%%%%%%%%%%%%%%%%%%%%%%%%%%%%%%%%%%%%%%%%%

\section*{Acknowledgements}
MR acknowledges funding from The Independent Research Fund Denmark, grant number DFF 1056-00027B. MR finalized the manuscript at $ \text{CP}^3$-Origins, University of Southern Denmark, as a postdoctoral researcher.

%%%%%%%%%%%%%%%%%%%%%%%%%%%%%%%%%%%%%%%%%%%%%%%%%%%%%%%%%%%%%%%%%%%%%%%%%%%%%%%%%%%%%%%%%%%%%%%%%%%%
%
\bibliography{CHvSIDM.bib}
\bibliographystyle{JHEP}
%
%%%%%%%%%%%%%%%%%%%%%%%%%%%%%%%%%%%%%%%%%%%%%%%%%%%%%%%%%%%%%%%%%%%%%%%%%%%%%%%%%%%%%%%%%%%%%%%%%%%% 

\end{document}